# Interpretable Machine Learning for Urban Heat Mitigation: Attribution and Weighting of Multi-Scale Drivers

**David Immanuel Tschan[1,2], Zhi Wang[1]; Dominik Strebel[1], Jan Carmeliet[1]; Yongling Zhao[1]**

[1]Department of Mechanical and Process Engineering, ETH Zürich, Switzerland

[2]Department of Environmental Systems Science, ETH Zürich, Switzerland

Corresponding author: Yongling Zhao (yozhao@ethz.ch)

**Key Points:**

- Machine learning allows characterising urban heat islands, accounting for drivers across scales with different controllabilities through urban planners
- Pre-classification of urban heat drivers into groups acting on the same scale and showing similar controllabilities allows mitigation-oriented feature ranking
- Adapting unit weight for all categories allows an optimal number of features from the highly controllable feature category to inform a mitigation strategy

## Abstract

Urban heat islands (UHIs) are often accentuated during heat waves (HWs) and pose a public health risk. Mitigating UHIs requires urban planners to first estimate how urban heat is influenced by different land use types (LUTs) and drivers across scales — from synoptic-scale climatic background processes to small-scale urban- and scale-bridging features. This study proposes to classify these drivers into driving (D), urban (U), and local (L) features, respectively. To increase interpretability and enhance computation efficiency, a LUT-distinguishing machine learning approach is proposed as a fast emulator for Weather Research and Forecasting model (WRF) coupled to the Noah land surface model (LSM) to predict ground- (TSK) and 2-meter air temperature (T2). Using random forest regression (RFR) with extreme gradient boosting (XGB) trained on WRF output over Zurich, Switzerland, during heatwave (HW) periods in 2017 and 2019, this study proposes LUT-based (LB) models that categorize features by scales and practical controllability, allowing optional categorical weighting. This approach enables category-specific feature ranking and sensitivity estimation of T2 and TSK to most important small-scale drivers — most notably surface emissivity, albedo, and leaf area index (LAI). Models employing the LB framework are statistically significantly more accurate than models that do not, with higher performance when more HW data is included in training. With RFR-XGB robustly performing optimal with unit weights, the method substantially increase interpretability. Despite the needs to reduce uncertainties and test the method on other cities, the proposed approach offers urban planners a direct framework for feasibility-centered UHI mitigation assessment.

## Plain Language Summary

Urban heat islands (UHIs) have major consequences for human health, especially during heat waves (HWs). To reduce temperatures in cities, knowing which features may lead to the highest temperature increases is important. Often, the Weather Research and Forecasting (WRF) model has been used for this. This study proposes a method to increase speed and interpretability of feature importance estimates. For this, random forest regression (RFR) and extreme gradient boosting (XGB) are used to emulate WRF coupled to the Noah land surface model (LSM) to predict ground- and 2m temperature (TSK, T2) over Zurich, Switzerland, for HW periods in 2017 and 2019. Before fitting, features are grouped into categories belonging to the different scales at which parameters influence UHIs, and categories may be weighted. Features within each category are then ranked and the most important features from the small-scale category are used to estimate mitigation potentials. Models using this method perform better than if the method is not applied, and predictions become more accurate if more HW data is used to train RFR-XGB. The method robustly indicates optimal weights of 1, and successfully identifies the most important small-scale drivers as emissivity, albedo and leaf-area index (LAI). More work is required to decrease uncertainties and demonstrate that the method works for other cities. Nevertheless, this study shows that the method produces highly interpretable and meaningful results, and that it offers a tool that urban planners can use to create UHI mitigation strategies with high feasibility.

**List of Acronyms**

| | |
|---|---|
| BUP | bulk urban parametrization |
| CB | city-based |
| CV | cross validation |
| D | driving |
| EMISS | emissivity |
| HP | hyperparameter |
| HW | heatwave |
| L | local |
| LAI | leaf area index |
| LB | land-based |
| LSM | land surface model |
| LUT | land use type |
| ML | machine learning |
| NHW | non-heatwave |
| NWP | numerical weather prediction |
| RFR | random forest regression |
| SBS | sequential backward selection |
| T2 | 2m-above ground temperature |
| TD | training data |
| TSK | surface temperature |
| U | urban |
| UA | urban area |
| UCM | urban canopy model |
| UHI | urban heat island |
| WD | working domain |
| WRF | Weather Research and Forecasting |
| XGB | extreme gradient boosting |

# 1 Introduction

Urban areas (UAs) characterized by higher air- and surface temperatures compared to surrounding rural areas are often called urban heat islands (UHIs) (Oke, 1982). Specific configurations in UAs lead to microclimates with surface temperature (TSK) and 2m-temperature (T2) up to $4 - 11$ °C (Hassan et al., 2021; Keith et al., 2023; Varentsov et al., 2018) and 7.8 °C (Phelan et al., 2015), respectively, warmer than in rural surroundings, driven through altered energy and humidity exchanges (F. Chen et al., 2014). Larger UAs associated with larger impermeable surfaces increase TSK (Hua et al., 2020) and tend to aggravate UHIs-caused heat stress (Ren et al., 2023; R. Wang, 2023), particularly during heat wave (HW) events (Ward et al., 2016) that enhance nighttime urban heating (Kong et al., 2023). These may increase in the future (Intergovernmental Panel On Climate Change (IPCC), 2023). Associated heat exposure effects include elevated energy consumption and water demand, deteriorating air quality and human health risks (Boned Fustel et al., 2021; Ebi et al., 2021; Ward et al., 2016) especially for groups at risk (D. Li & Bou-Zeid, 2013; Verein Klimaseniorinnen Schweiz and Others v. Switzerland, 2024; Yang et al., 2021). With UAs likely to grow and host larger shares of populations (United Nations Organisation, 2019), UHIs are a major effect of climate change and urbanisation and pose an increasing public health risk.

Minimising UHI impacts requires implementation of heat mitigation strategies. To inform on the applicability of such strategies, numerical weather prediction (NWP) systems on synoptic- to meso scales have proven effective in assessing UHI drivers. Frequently, the Weather Research and Forecasting (WRF) model (Skamarock et al., 2021) is employed to this end, demonstrating good performance under heatwave (HW)- and non-heatwave (NHW) conditions (F. Chen et al., 2014; Giannaros et al., 2018; H. Li et al., 2019; Patel et al., 2022). WRF may be used in conjunction with urban canopy models (UCMs) and bulk urban parametrizations (BUPs), and it is generally applied using both although several studies omit an UCM. WRF has been used to assess mitigation strategies such as increased urban vegetation and cooling roofs (Cui & Foy, 2012; X.-X. Li & Norford, 2016), and to highlight major UHI drivers across scales such as meso- or synoptic-scale atmospheric circulation (Aquino-Martínez et al., 2025), urban scale surface albedo and emissivity configurations (Giannaros et al., 2018), urban anthropogenic heat release (F. Chen et al., 2014), or scale-bridging parameters such as nighttime urban boundary layer structure (Cui & Foy, 2012). Such studies also indicate that UHIs may vary with terrestrial morphology, such as proximity to oceans and elevation (Vahmani & Ban-Weiss, 2016), background climate (L. Zhao et al., 2014) (using the Community Earth System Model), and urban land-use types (LUTs) (F. Chen et al., 2014; X.-X. Li & Norford, 2016; Patel et al., 2022). Consequently, with drivers varying across scales, heat mitigation strategies based on driver characterization are expected to have different effects in different cities (Georgescu et al., 2014).

These findings emphasise the need to design city-specific UHI mitigation strategies that account for processes across all scales and LUT-induced UHI variations. Importantly, different drivers belonging to different scales may exhibit different degrees of modifiability (Y. Zhao et al., 2025). This, along with the fact that different LUTs may be associated with different drivers, may lead to difficulties for urban planners to (1) assess which scales are most important for their UAs; (2) determine which features to choose for mitigation measures out of the pool of modifiable features within their UAs; and (3) estimate the resulting mitigation potentials given the scale at which the chosen features affect UHIs. Once a pool of measures is known, recently frameworks to decide on particular mitigation strategies have been developed (Y. Zhao et al., 2023). However, urban planners face a lack of effective methodology to systematically construct such a pool, based on efficient assessment of mitigation potentials for their respective cities and local contexts (Carmeliet & Derome, 2024).

To alleviate this issue, NWP-based schemes provide a viable tool. Recent work suggests that NWP schemes adopting BUPs but omitting UCMs require less computational resources than full NWP-BUP-UCM models, and may perform reasonably well for urban heat mitigation assessment (Y. Zhao et al., 2024). Nevertheless, NWP applications for UHI research remain challenging and computationally expensive (Andraju et al., 2019; F. Chen et al., 2014; Lean et al., 2024; J. Wang et al., 2019; Zhong et al., 2023). For mitigation assessment, supervised machine learning (ML) is an attractive alternative to fully or partially emulate NWP schemes, delivering fast estimators for mitigation potentials. A promising method is random forest regression (RFR) employing extreme gradient boosting (XGB). It is efficient with large datasets, requires limited hyperparameter (HP) tuning and offers high physical interpretability. RFR-XGB has been used to identify seasonal differences in UHI drivers (Liu et al., 2023) and to suggest urban greening and permeable surfaces as UHI mitigation strategies (Mohammad et al., 2022). The results of well-calibrated RFR-XGB models are largely in line with NWP scheme predictions, and they may therefore provide viable alternatives or additions to NWP-based systems. RFR-XGB has also enabled measurement of the impact of different indices related to water and vegetation in UAs on UHIs (Garzón et al., 2021), and has been used to assess the importance of different LUTs on UA land surface temperatures (McCarty et al., 2021). These are important steps towards assessing the importances of urban climate drivers to inform pertaining mitigation potentials in a computationally efficient and fast way. Nevertheless, these results are constrained to particular cities or climatic regions and offer limited generalisability into a universal framework.

This study aims to close that gap by first introducing a ML method tailored for UHI mitigation. In addition, it proposes, examines, and demonstrates the critical important of ranking and selecting features with a similar degree of modifiability for UHI mitigation — both when distinguishing LUTs in the model and when not, with such distinguishing accounting for the different physical regimes in surface energy balance across LUTs found in various studies (F. Chen et al., 2014; X.-X. Li & Norford, 2016). It is assumed that small-scale drivers are generally more easily modifiable, whereas larger scale drivers are harder to modify or not modifiable at all, which allows focusing on relevant, modifiable drivers for mitigation potential assessment. To this end, prior to ranking and selecting optimal feature sets, features are pooled into categories that impact UHIs on similar scales and exhibit similar degrees of controllability by urban planners. Then, optimal feature sets are subjected to differential weighting across categories to determine the relative impact of different scales on UHIs. To obtain feature importances and mitigation potentials, RFR-XGB regression is used to emulate the WRF model coupled to Noah land surface model (LSM) to predict surface- and 2m-air temperatures (TSK, T2), respectively.

Specifically, the aims of the study are (1) to show that HW data is required in the training data set to reproduce HW conditions in predictions; (2) to demonstrate that LUT-specific land-based (LB) models increase performance compared to city-based (CB) models that do not distinguish LUTs; and (3) to show that LB models applying the proposed method — feature pre-classification, subsequent feature set optimization, and weighing of features across categories — maximise both interpretability and performance. This is demonstrated for Zurich, Switzerland, where a significant UHI has been reported (Canton & Dipankar, 2024). While the demonstration is limited to a single UA, the approach is designed to be generalisable and can be extended to urban heat mitigation studies in cities with diverse local contexts. Thus, this study provides decision-makers and urban planners with a scalable method to assess the dominant UHI-driving scale in their urban area, to select the most relevant modifiable features for mitigation, and quantify the associated mitigation potentials.

## 2 Method

### 2.1 Overview

This study uses output from Advanced Research WRF version 4.2 (Skamarock et al., 2021) to train a RFR-XGB emulator for T2 and TSK. Y. Zhao et al. (2024) describe the model configuration used in more detail. Importantly, no UCM was used, but Noah LSM and a BUP was adopted. WRF was run without down-nesting at a resolution of $1km \times 1km$ using a spin-up period of $72h$. Spatial data extent of the model output thus is $309km$ west-east and $189km$ south-north over middle Europe (fig. 1ab).

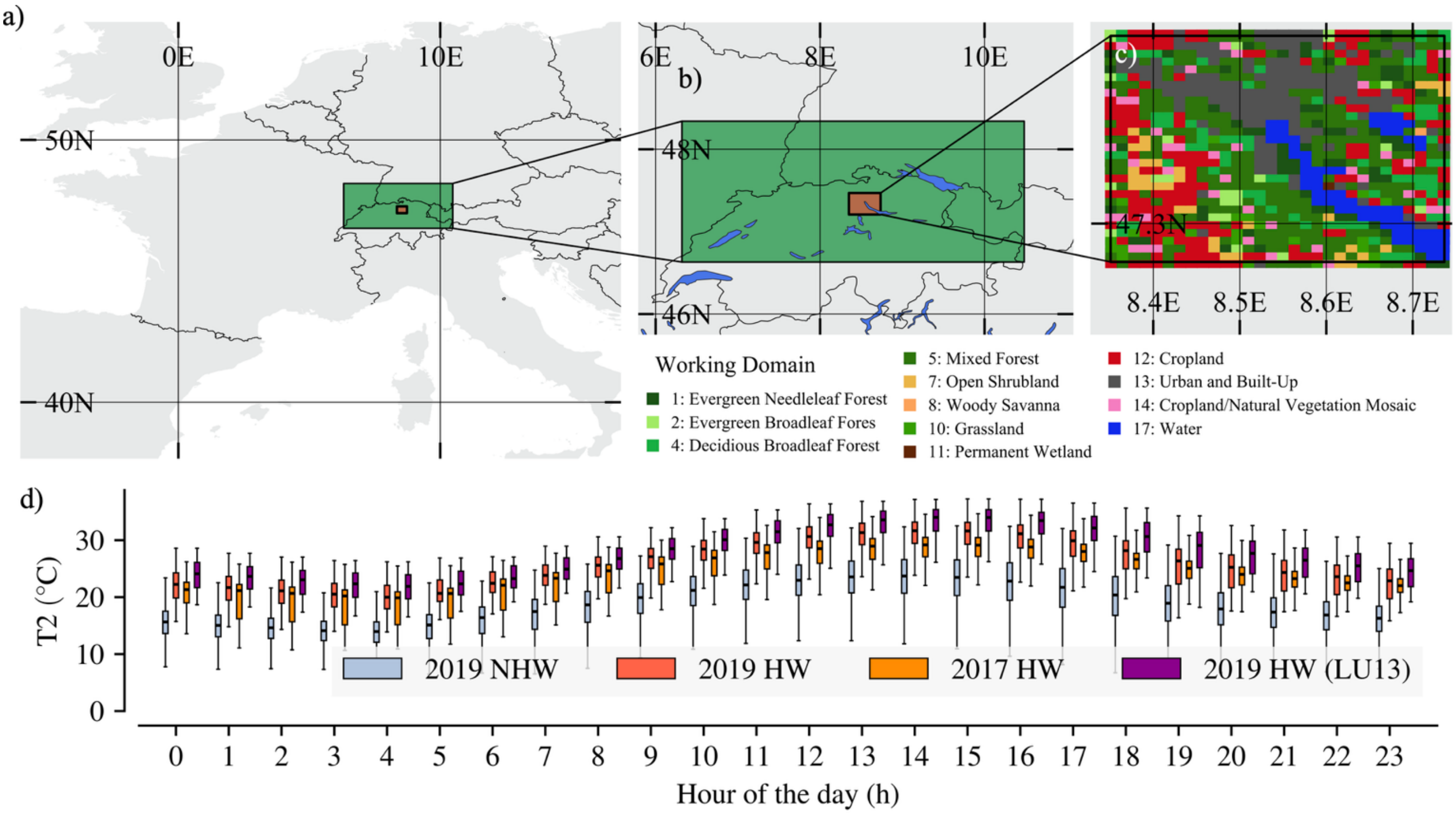

Figure 1: a: Overview over domains within Europe. b: WRF domain (green) and working domain (WD) (red) over Switzerland. c: Detailed view of WD over Zurich, Switzerland indicating LUTs. d: Diurnal cycle of T2 for the WD for different data sets. There is a clear distinction in the diurnal cycle for HW and NHW conditions, with air temperature highest during HW for LUT 13. Indicated are median temperatures, while bars extend to 25th and 75th percentiles, respectively. Shown values pertain to WD without including LUT 17 (water).

From this, a subspace spanning $30km \times 30km$ over the region of Zurich, Switzerland (fig. 1c) is selected as the study working domain (WD). Temporally, data covers two periods, July 2017 including only HW data, and June and July 2019 containing both HW and NHW data. HW data show higher T2 over the diurnal cycle than NHW data (cf. fig. 1d), especially pronounced for LUT 13 (urban and built-up). The WRF data contains 24 variables (see table S1, supplementary materials), with T2 and TSK used as targets and all others as features, and 12 different LUTs following Sulla-Menashe & Friedl (2018) (see table S2, supplementary materials). Data with LUT representing water (LUT 17) was removed for both CB- and LB-models because water areas follow different energy balance regimes than terrestrial LUTs.

## 2.2 City-Based Models and Land-Based Models

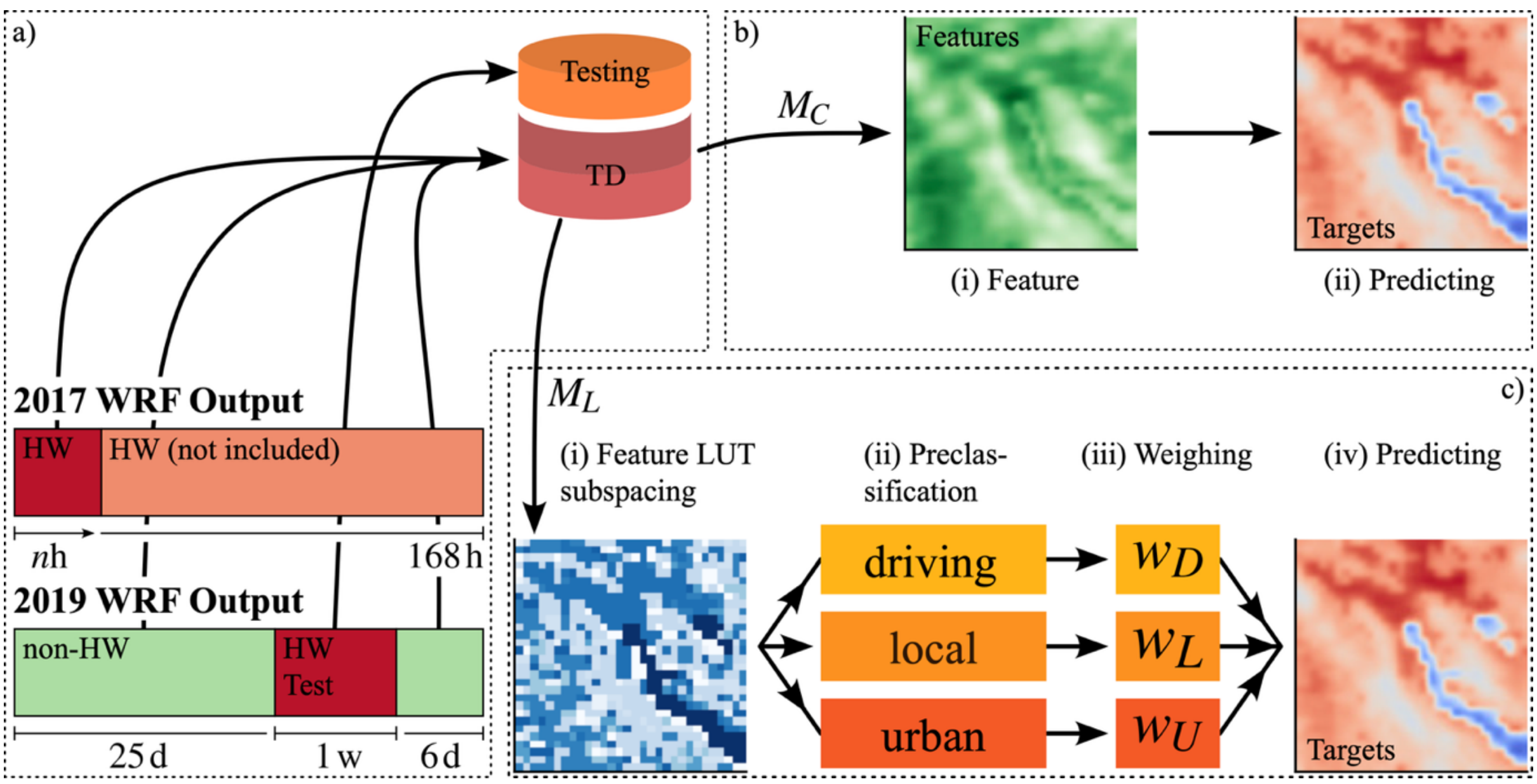

Figure 2: a: Aggregation of data into a dataset. 2017 heat wave (HW) data and 2019 non-heat wave (NHW) data are aggregated into the training data (TD), while the test data consists of 2019 HW data. 2017 HW training data may be varied between 0 and 168 (the maximum) included data hours. b: City-based (CB) model $M_C$. In such models, features from the dataset (i) are directly used to train RFR-XGB to predict T2 and TSK (ii). c: Land-based (LB) model $M_L$. In such models, the feature space is first split into LUT subspaces (i). The features are then pre-classified into driving (D)-, local (L) and urban (U)-features that pertain to synoptic-scale-, scale-bridging- and urban-scale processes (ii). The features in each group are then subjected to weighting by multiplying all features from each category with the pertaining weight $w_D$, $w_L$, $w_D$ (iii). Finally, a CB model is trained on this LUT-specific data to predict T2 and TSK for the pertaining LUT (iv). The complete LB model consists of the aggregated CB models, each pertaining to a specific LUT.

### 2.2.1 City-Based Models

For city-based (CB) models, data is aggregated from WRF output (fig. 2a). All 2019 NHW data as well as variable number of 2017 HW data hours are included in the training set, and the testing set consists of all 2019 HW data. CB models (fig. 2b) do not distinguish LUTs. Instead, a CB model including $n_t$ features, denoted $M_C^{(n_t)}$, learns the relationship of features — independently of the scale they pertain to or their degree of modifiability in a mitigation setting (fig. 2bi) — and T2 and TSK and predicts estimates of these targets (fig. 2bii).

### 2.2.2 Land-Based Models

For land-based (LB) models, data is aggregated similarly to CB models (fig. 2a). Unlike CB models, however, LB models (fig. 2c) distinguish LUTs to account for different physical energy balance regimes, classify features into driving (D)-, local (L) and urban (U) features — pertaining to meso- to synoptic scales, scale-bridging scales and small urban scales — and assign weights to them. Classification accounts for different underlying scales and associated differences in modifiability, while weighting represents the differential influence of scales on UHIs. Specifically, a LB model containing $n_D$ (D)-, $n_L$ (L)- and $n_U$ (U)-features, denoted $M_L^{(n_D,n_L,n_U)}$, splits the feature space into LUT specific subspaces (fig. 2ci) and fits CB submodels on each LUT subspace, such that the LB models consists of the totality of CB submodels. After LUT-splitting, features are pre-classified into categories (fig. 2cii) and each category may be weighted with weights $w_D$, $w_L$, $w_U$ to represent the differential importance of D-, L- and U-scales (fig. 2ciii). The model $M_L^{(n_D,n_L,n_U)}$ then learns the relationship between features and targets and predicts an estimate of T2 and TSK (fig. 2civ). To obtain TSK and T2 predictions across the entire WD, the estimates must be aggregated across LUT-specific submodels.

## 2.3 Data, Preprocessing and Feature Categorization

### 2.3.1 Data

The WRF data used to train CB- and LB models has spatial resolution of $1km$ and temporal resolution of $1h$. HW data for 2017 spans the period between 18.06.2017 and 24.06.2017. NHW data of 2019 spans the period between 01.06.2019 to 08.07.2019, except for a HW event between 25.06.2019 and 02.07.2019. Totally, 2017 data includes $168h$ of HW data, and 2019 data includes 25 days of NHW data followed by $1w$ of HW data followed by 6 days of NHW data. The training dataset (TD) for RFR-XGB consists of up to 168 variable hourly data slices of 2017 HW data and the entire 2019 NHW data (fig. 2a). The test dataset consists of the week of HW data from 2019. Because it was assumed that any NHW sample should be an element of the HW sample space as well, no NHW data was included in the testing.

### 2.3.2 Preprocessing

During preprocessing, data is scaled and shuffled. Scaling is implemented to assure all features are of similar order of magnitude so that RFR-XGB does not learn on features' absolute magnitudes. Because feature distributions are different (see table S1, supplementary material: consider mean and skewness of features), data was scaled using the min-max approach (Tanoori et al., 2024). Importantly, training and testing features must be scaled with the same scaling obtained from the TD only. After scaling, data is randomly shuffled to reduce overfitting. There is a set of categorical features present in the data, including soil type, vegetation type and, for CB models, LUT. Since these are present on an ordinal scale, they were subjected to the same preprocessing steps as numerical features and not treated specially.

### 2.3.3 Feature Pre-Classification

Features were pre-classified into the U- (urban or small-scale), L- (local or scale-bridging), and D-features (driving or meso- to synoptic scale) in LB models (see table S1, supplementary materials). Pre-classification thereby accounts for both the assumed scale-modifiability relationship and the notion that feature ranking is most meaningful for mitigation potential assessment if the ranked features are modifiable to the same degree. For example, solar zenith angle or terrain height are driving features, whereas leaf area index (LAI) or vegetation type are urban features. While solar zenith angle and terrain height are not modifiable, LAI or vegetation types in cities are modifiable. Low level wind speed is a local feature, as it is determined both by synoptic-scale atmospheric states and small-scale urban 3D structure. Accordingly, its degree of modifiability is intermediate: altered urban building morphology — an aspect that may be controlled in urban design — may allow for more circulation. However, circulation is subject to large-scale controls through synoptic-scale circulation or proximity to oceans or topography — aspects not controllable in cities.

Detailed justification why any given feature belongs in any given category can be found in the supplementary materials (see text S1). Importantly, categorization is dependent on the available data and the UA for which it is applied. From the total of $N_t = 22$ features available for CB models, there are $N_D = 6$ D-features, $N_L = 10$ L-features and $N_U = 5$ D-features considered in this study, with LUT not considered a feature in LB models.

### 2.4 Metrics

The metrics used in this study are the coefficient of determination $R^2$, the root mean squared error RMSE, the mean absolute error MAE and the mean bias MB (see text S2, supplementary materials). The critical metric used for model selection was $R^2$, that is $R^2$-scores were used for HP tuning, feature ranking, feature set optimization, and model selection. Let X denote any of the used metrics. Then:

1. $X_{avg}$ refers to the average score across the two targets TSK and T2: a separate model was fitted for each target and $X_{avg}$ is the average accuracy across the two targets.
2. ${}_{s}X$ refers to scores obtained using identical random shuffling.
3. $\mu(X\{\mathcal{N}_k\})$ refers to the mean of scores obtained from an ensemble $\mathcal{N}_k$ of scores.

### 2.5 Experimental Workflow

This study uses a sequential methodology:

1. Fitting baseline CB and LB models $M_{C,0}^{(22)}$, $M_{L,0}^{(6,10,5)}$, respectively, that include all features. In this step HPs are tuned using 10-fold cross validation (CV) and all $168h$ of 2017 HW data. For LB models, this means that the underlying set of CB models is subjected to HP tuning, such that HPs differ across LUTs. Subsequent models use the same HPs as the baseline models, whereas all other parameters — particularly the number of features and HW training hours form 2017 — may be varied. Data aggregation follows sec. 2.3.1 and fig. 2, respectively.
2. Running a feature ranking scheme based on sequential backward selection (SBS) to obtain feature-inclusion orders (FIOs) (see sec. 2.7) that represent a feature importance ranking. For LB models, FIOs are LUT- and category-specific.
3. Optimizing feature sets — that is finding optimal $n_t$ for CB models and optimal $(n_D, n_L, n_U)$ for LB models — using variable numbers of features from the FIOs. This results in optimised CB and LB models denoted with $M_{C,B}$, $M_{L,B}$, respectively.
4. Applying the resulting optimal models to assess the mitigation potential of the most important features identified in step 3.

### 2.6 Random Forests and Extreme Gradient Boosting

This study uses RFR-XGB, where XGB is provided via the open-source XGBoost library developed by T. Chen & Guestrin (2016). RFR-XGB may result in estimators with low variance and high predictive power, in a computationally efficient manner that may be fitted on a single personal computer. While other ML methods, particularly neural networks, may be well-suited for UHI mitigation research (H. Wang et al., 2023), associated computational costs and required data quantities are high and the interpretability low.

### 2.7 Feature Ranking Scheme

Once the baseline CB and LB models ($M_{C,0}$ and $M_{L,0}$) have been established, a feature selection scheme is proposed and implemented, based on SBS (Raschka & Mirjalili, 2019). Contrastingly to regular SBS implementations, this study identifies the feature causing the

most — rather than the least — decrease in performance at each iteration and removes it from the feature set. While in CB models, this feature selection scheme is applied to the entire feature set, it needs to be applied for each LUT submodel and for each of the D-, L- and U feature categories separately in LB models. Let $F^{(n)}$ denote the baseline feature set consisting of $n$ features, that is, $F^{(n)} = \{f_1, f_2, \dots, f_n\}$. For CB models, this is a single feature set $F^{(n_t)}$ containing all $n_t = 22$ features while for LB models, there are three baseline feature sets $F^{(n_D)}$, $F^{(n_L)}$, $F^{(n_U)}$ with $n_D = 6$, $n_L = 10$, $n_U = 5$ features, pertaining to D-, L-, and U-features, respectively. For $F^{(n_t)}$, $F^{(n_D)}$, $F^{(n_L)}$ and $F^{(n_U)}$ the feature selection scheme starts at iteration $t = 0$ and recursively creates a list ${}_*F$, referred to as a feature inclusion order (FIO) (see fig. 3).

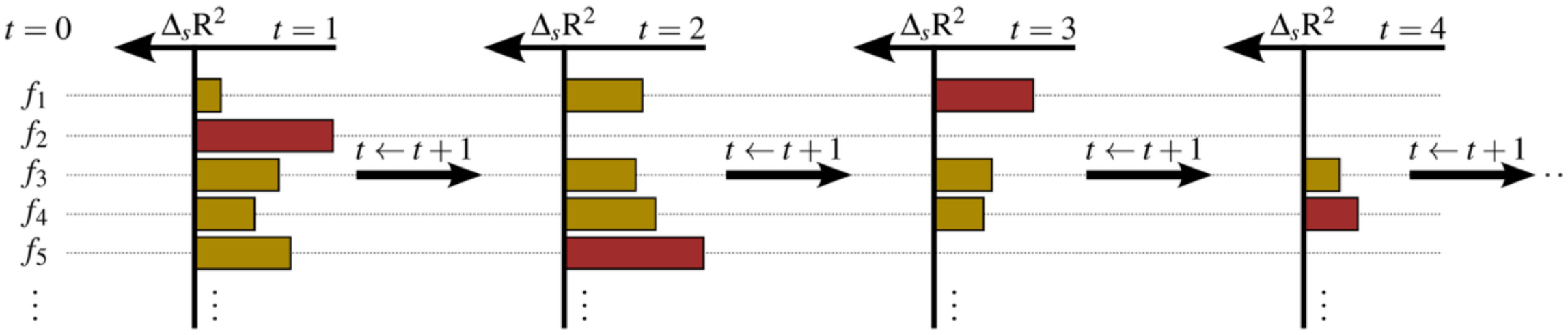

Figure 3: Algorithm to create feature inclusion orders (FIOs). In each iteration (denoted as $t$), the feature whose removal from the feature set causes the greatest loss in accuracy (red bar) is identified and removed in the subsequent iteration. That feature is then recorded in the FIO. When the algorithm terminates, it has created an ordered set of features from the initial feature set, denoted ${}_*F$.

In each iteration $t$, the scheme identifies the most important feature ${}_*f$ as the one causing the largest decrease in ${}_s\mathrm{R}^2_{\mathrm{avg}}$ when left out. This feature is then recorded at position $t$ of the FIO (for a more detailed outline see text S3, supplementary materials).

Importantly, FIOs do not results in an absolute importance scale for features. Given the FIO ${}_*F = \{{}_*f_i\}_{i=1}^{n}$, claiming that ${}_*f_k$ is more important than ${}_*f_j$ if $j < k \leq n$ holds not *per se*, because ${}_*f_k$ was determined from a feature set from which ${}_*f_j$ has already been removed. Rather, FIOs represent a *hierarchical* importance scale: given the FIO ${}_*F = \{{}_*f_i\}_{i=1}^{n}$, for $s < n$, if feature ${}_*f_s$ is included, the next feature to be included should be ${}_*f_{s+1}$ in order to cause the highest increase in variability explicable by the model. Feature importance in this study should be understood in this hierarchical sense. As the proposed scheme is based on a greedy method, a globally optimal feature set may be missed.

## 2.8 Optimal Feature Sets and Weighting of Categories

Through increasing the number of ordered features, FIOs may serve to construct sequential feature sets $F^{({}_*m)} = \{{}_*f_i\}_{i=1}^{{}_*m \leq n}$. By finding $F^{({}_*m)} = \arg\max_{{}_*k} \mathrm{R}^2_{\mathrm{avg}}\left(F^{({}_*k)}\right)$, an optimal feature set may be identified. For CB models, the identified optimal feature set with ${}_*n_t$ features pertains to the entire WD and allows to find an optimal model $M_{C,B}^{({}_*n_t)}$.

For LB models, per LUT, three FIOs — pertaining to D-, L- and U-features — are required, to find the optimal numbers of features ${}_{*}n_{D}$, ${}_{*}n_{L}$, ${}_{*}n_{U}$, which may then be aggregated into an optimal model $M_{L,B}^{({}_{*}n_{D},\ {}_{*}n_{L},\ {}_{*}n_{U})}$. This requires including the same number of features per category across LUTs, even though the included features per category are different across LUTs. Such models imply unit weight of all categories and are hence called *non-weighted*, denoted ${}_{nw}M_{L,B}^{({}_{*}n_{D},\ {}_{*}n_{L},\ {}_{*}n_{U})}$.

Contrastingly, weights $w_D, w_L, w_U$ for driving-, local-, and urban- features differing from unity may be assigned to each category, resulting in models referred to as *weighted*, denoted ${}_{w}M_{L,B}^{({}_{*}n_{D},\ {}_{*}n_{L},\ {}_{*}n_{U})}$. Importantly, these weights are *not* computed internally by RFR-XGB but are *assigned* to the categories to reflect their relative importance, and are hence different from the weights normally associated with RFR. In this study, *weights* explicitly refer to these *assigned* weights. Detailed methodology on how to obtain such weights for categories is outlined in the supplementary materials (see text S4).

### 2.9 Procedure to obtain Optimal Feature Sets

Through sequentially including more features from the FIO, a locally optimal CB model $M_{C,B}$ may be identified. To find the best non-weighted LB model ${}_{nw}M_{L,B}$, the total combinatorial search space of $N_D \cdot N_L \cdot N_U = 300$ feature permutations was tested, from which the top five permutations were chosen and ran in weighted mode to produce a total of ten optimised LB models. Finally, another optimised city model was developed, constrained to include no more features than the optimised LB model with the highest total feature count. In total, this results in an array of twelve optimised models along with two baseline models.

### 2.10 Model Variance Estimation and Statistical Significance Testing

#### 2.10.1 Model Variance Estimation

The feature optimization scheme results in a set of models whose performance is at least locally optimal. However, all models trained so far have been subjected to the same random seed for shuffling in the data preprocessing steps. This means that (1) inter-model variability based on ${}_{s}\mathrm{R}^2_{\mathrm{avg}}$ scores is likely due to different features and processing (LB vs. CB); and (2) ${}_{s}\mathrm{R}^2_{\mathrm{avg}}$ scores may not represent scores the model would deliver upon shuffling with different random seeds. To address this, each model was fitted onto the data $\epsilon_1$ times with different random seeds, resulting in accuracy distributions $\mathrm{R}^2\{\mathcal{N}_{i,\epsilon_1}\}$ where $\mathcal{N}_{i,\epsilon_1}$ denotes the distribution of the $\epsilon_1$ runs associated with model $i$. Here, $\epsilon_1 = 200$ was chosen arbitrarily, as this value is above minimal required sample size for the performed statistical test (Kim & Park, 2019). The expectation values of the distributions $\mu\left(\mathrm{R}^2_{\mathrm{avg}}\{\mathcal{N}_{i,\epsilon_1}\}\right)$ are more robust measures of model accuracies than the ${}_{s}\mathrm{R}^2_{\mathrm{avg}}$ scores. Through obtaining variance and skewness estimates, the distributions furthermore help to quantify the variability that may be expected from the models. The ensemble score $\mu\left(\mathrm{X}_{\mathrm{avg}}\{\mathcal{N}_{i,\epsilon_1}\}\right)$ was also computed for the other metrics used.

#### 2.10.2 Statistical Significance and Normality Testing

For any pair of models $M_i$, $M_j$ with accuracy distributions $\mathrm{R}^2_{\mathrm{avg}}\{\mathcal{N}_{i,\epsilon_1}\}$, $\mathrm{R}^2_{\mathrm{avg}}\{\mathcal{N}_{j,\epsilon_1}\}$, a paired $t$-test is performed in which the two distributions are tested against the null hypothesis of having identical means. The difference of means $\Delta\mu_{ij} \equiv \mu\left(\mathrm{R}^2_{\mathrm{avg}}\{\mathcal{N}_{i,\epsilon_1}\}\right) - \mu\left(\mathrm{R}^2_{\mathrm{avg}}\{\mathcal{N}_{j,\epsilon_1}\}\right)$ is considered statistically significant if the $p$-value associated with the $t$-test is $p \leq 0.005$ (Benjamin et al., 2018). Using $p$-values, any model $M_i$ may be furthermore tested against the null hypothesis that its accuracy distribution is normal. Statistical significance again requires $p < 0.005$.

### 2.11 Model Application for Mitigation Potential Estimation

To assess the mitigation potential associated with the ${}_*n_U$ U-features over UAs — for LUT 13 (urban and built-up), that is — the testing data of each feature ${}_*f_k$ included in the optimal U-feature set for LUT 13 $F^{({}_*n_U)} = \{{}_*f_k\}_{k=1}^{*n}$ may be increased or decreased by a value ${}^{\uparrow}\Delta_k$ or ${}_{\downarrow}\Delta_k$, respectively. Both altering a single U-features or varying several U-features simultaneously is possible. The change in the target value $\Delta\widehat{\mathrm{T}}_i$ for model $i$ and target T — either T2 or TSK — is then estimated as the difference between average predictions from non-varied and varied U-features. From this, a target sensitivity $\Delta\widehat{\mathrm{T}}_i\ \delta^{-1}_{*f_k}$ towards unit change of ${}_*f_k$, that is $\delta^{-1}_{*f_k}$, may be estimated via

$$\Delta\widehat{\mathrm{T}}_i\ \delta^{-1}_{*f_k} = \frac{\partial \Delta\widehat{\mathrm{T}}_i}{\partial\ {}_*f_k} \approx \frac{\Delta\widehat{\mathrm{T}}_i\left({}_*f_k + {}^{\uparrow}\Delta_k\right) - \Delta\widehat{\mathrm{T}}_i\left({}_*f_k - {}_{\downarrow}\Delta_k\right)}{{}^{\uparrow}\Delta_k - {}_{\downarrow}\Delta_k} \tag{1}$$

The average prediction for varied U-features is obtained through running an ensemble of model $i$ comprising $\epsilon_2 = 200$ runs, resulting in model distributions $\mathcal{N}_{i,\epsilon_2=200}$. To account for different behaviour between day periods (06:00 – 21:59) and night periods (22:00 – 05:59) (F. Chen et al., 2014), these may be treated separately, but a total response across the full diurnal cycle may be computed as well. A detailed overview over the methodology is available in the supplementary materials (see text S5).

## 3 Results

The results of this study may be structured into four subsections. The first one will describe the results of the baseline models — based on ${}_s\mathrm{R}^2_{\mathrm{avg}}$ scores — and how these were used to find optimal feature sets. Thereafter, a statistical analysis — pertaining to $\mathrm{R}^2_{\mathrm{avg}}\{\mathcal{N}_{i,\epsilon_1}\}$ scores — is provided in which expected model variance and optimal model selection is presented. The subsequent subsection outlines how selected optimal models behave under different number of included HW data hours, while the final subsection consists of results of applying the selected optimal models to estimate mitigation potentials.

### 3.1 Baseline Models, Feature Selection and Feature Set Optimization

The baseline models result in ${}_s\mathrm{R}^2_{\mathrm{avg}}$ accuracies of 0.8163 and 0.8084 for $M_{C,0}$ and $M_{L,0}$, respectively, with the CB model having higher accuracy (lower ${}_s\mathrm{RMSE}_{\mathrm{avg}}$ and ${}_s\mathrm{MAE}_{\mathrm{avg}}$) and lower bias (lower ${}_s\mathrm{MB}_{\mathrm{avg}}$) than CB models (all metrics listed in table S3, supplementary materials). Both models have high negative bias ($-1.45 \ldots -1.3$ °C) and relatively high ${}_s\mathrm{RMSE}_{\mathrm{avg}}$ (2.28 ... 2.36 °C) and ${}_s\mathrm{MAE}_{\mathrm{avg}}$ (1.76 ... 1.83 °C).

Employing these baseline models and using the methodology proposed in sec. 2.7, FIOs for the CB and LB models were found (see tables S5 and S5, supplementary materials, respectively) and used to find optimal feature sets using the approach outline in sec. 2.9. For CB models, a relationship of test ${}_s\mathrm{R}^2_{\mathrm{avg}}$ with changing numbers of included features $n_t$ may be obtained (fig. 4abc). Fig. 4a shows that ${}_{s}\mathrm{R}^2_{\mathrm{avg}}$ is very low for $n_t = 1$ but increases rapidly. A local optimum is reached at $n_t = 7$ ($M^{(7)}_{C,B}$), and after a slight performance dip a global maximum at $n_t = 19$ ($M^{(19)}_{C,B}$) is reached. This behaviour is similar for T2 (fig. 4b) and TSK (fig. 4c) and is in line with reported supervised feature selection behaviour (Cai et al., 2018).

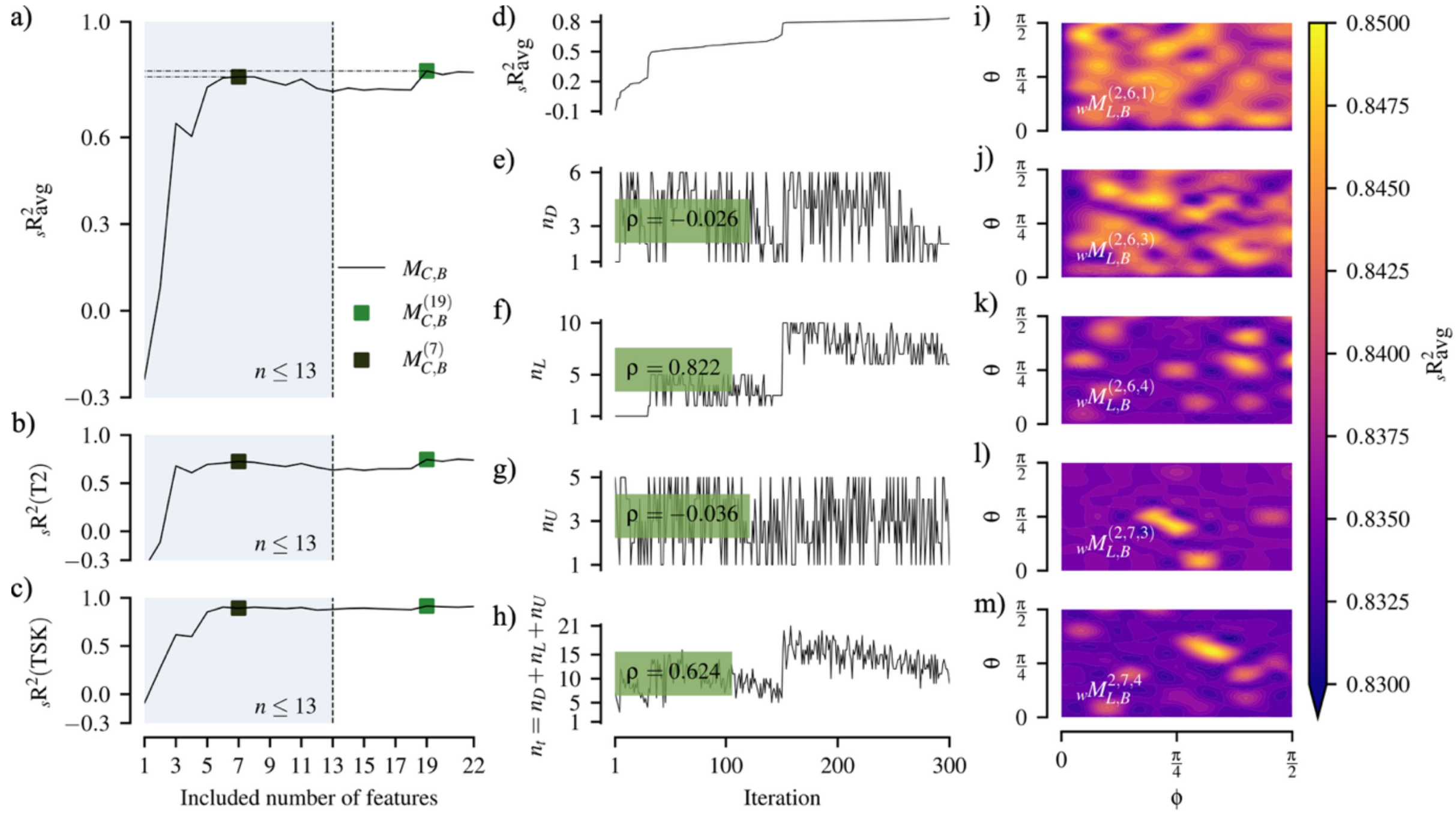

Figure 4: a – c: the change in accuracy when varying the number of included features from the FIO for the CB model. The baseline model $M_{C,0}$ herein refers to the very last model with 22 features included. d – h: the change in accuracy when varying the number of included driving-, local- and urban features for the LB non-weighted model, ordered here in ascending accuracy. Note that correlations between the number of included features and the accuracy is reported. Iteration here refers to a particular combination $(n_D, n_L, n_U)$. i – m: the variation of accuracy across $\phi, \theta$-space (see text S4, supplementary materials) of the top 5 non-weighted models from panel B, i.e. representing the weighted models. The displayed values stem from a $11 \times 11$ grid that has been interpolated using cubic interpolation. Models decrease in accuracy from top down.

Likely, too few features contain insufficient signal for RFR-XGB to learn targets, while too many features may cause overfitting to unimportant features.

Fig. 4d shows the accuracy of the LB model when testing all 300 possible permutations of included driving-, local- and urban- features, displayed in order of ascending ${}_{s}\mathrm{R}^2_{\mathrm{avg}}$. Figures 4e-h show the pertaining number of D-, L-, U-features as well as $n_t$. The ${}_{s}\mathrm{R}^2_{\mathrm{avg}}$ curve in fig. 4d has two discontinuities at about iterations 40 and 150 that seem to coincide with discontinuities in $n_L$ (local features). As furthermore suggested by the highest correlations coefficients — displayed as $\rho$ on figs. 4e-h and indicating the correlation coefficient between $n_D$, $n_L$, $n_U$ or $n_t$ and ${}_{s}\mathrm{R}^2_{\mathrm{avg}}$, respectively — of local features with ${}_{s}\mathrm{R}^2_{\mathrm{avg}}$, this is strong evidence that L-features may have high impact and are thus critical for building reliable urban climate ML models. This is plausible as L-features link local urban climatic factors with large-scale patterns and thus likely include critical features for urban temperatures. Ideally, the proposed methodology should reflect such importance in category weights. To determine these, the top five permutations in figs. 4e-h, ${}_{nw}M_{L,B}^{2,6,1}$, ${}_{nw}M_{L,B}^{2,6,3}$, ${}_{nw}M_{L,B}^{2,6,4}$, ${}_{nw}M_{L,B}^{2,7,3}$ and ${}_{nw}M_{L,B}^{2,7,4}$ ( ${}_{s}\mathrm{R}_{\mathrm{avg}}^{2} = 0.83 \ldots 0.84$, ${}_{s}\mathrm{RMSE}_{\mathrm{avg}} = 2.16 \ldots 2.22\ ^{\circ}\mathrm{C}$, ${}_{s}\mathrm{MAE}_{\mathrm{avg}} = 1.64 \ldots 1.69\ ^{\circ}\mathrm{C}$, $s\mathrm{MB}_{\mathrm{avg}} = -1.28 \ldots -1.14\ ^{\circ}\mathrm{C}$) were selected to be run in weighted mode.

The ${}_{s}\mathrm{R}_{\mathrm{avg}}^{2}$ accuracies of the pertaining weighted models ${}_{w}M_{L,B}^{2,6,1}$, ${}_{w}M_{L,B}^{2,6,3}$, ${}_{w}M_{L,B}^{2,6,4}$, ${}_{w}M_{L,B}^{2,7,3}$ and ${}_{w}M_{L,B}^{2,7,4}$ ( ${}_{s}\mathrm{R}_{\mathrm{avg}}^{2} = 0.83 \ldots 0.85$, ${}_{s}\mathrm{RMSE}_{\mathrm{avg}} = 2.13 \ldots 2.22\ ^{\circ}\mathrm{C}$, ${}_{s}\mathrm{MAE}_{\mathrm{avg}} = 1.62 \ldots 1.69\ ^{\circ}\mathrm{C}$, $s\mathrm{MB}_{\mathrm{avg}} = -1.27 \ldots -1.11\ ^{\circ}\mathrm{C}$) in function of $\phi - \theta$ space (see text S4, supplementary materials) are shown in figs. 4i-m and are slightly higher than their non-weighted counterparts. Across the ${}_{s}\mathrm{R}_{\mathrm{avg}}^{2}$- and ${}_{s}\mathrm{RMSE}_{\mathrm{avg}}$-metrics, weighted models slightly outperform their non-weighted counterparts. Similarly, ${}_{s}\mathrm{MAE}_{\mathrm{avg}}$ and ${}_{s}\mathrm{MB}_{\mathrm{avg}}$ are generally lower for weighted than for non-weighted models, although differences are small and ${}_{s}\mathrm{MB}_{\mathrm{avg}}$ scores still indicate a high negative bias, and ${}_{s}\mathrm{RMSE}_{\mathrm{avg}}$ and ${}_{s}\mathrm{MB}_{\mathrm{avg}}$ remain high. Furthermore, "better" (higher ${}_{s}\mathrm{R}_{\mathrm{avg}}^{2}$) models like ${}_{w}M_{L,B}^{2,6,1}$, ${}_{w}M_{L,B}^{2,6,3}$ (figs. 4ij) appear less sensitive to weighting than "worse" (lower ${}_{s}\mathrm{R}_{\mathrm{avg}}^{2}$) models like ${}_{w}M_{L,B}^{2,7,3}$, ${}_{w}M_{L,B}^{2,7,4}$ (fig. 4lm) that exhibit discernible ${}_{s}\mathrm{R}_{\mathrm{avg}}^{2}$ peaks towards $\phi = \theta = \pi/4$. These findings indicate that (1) "worse" models may perform well and almost equivalate "better" models when an appropriate set of weights is found and adopted; or (2) that "better" models are inherently able to predict the targets well, even when non-optimal weights are adopted. The weights pertaining to the "worse" models are approximately $w_D = w_L = 0.5$ and $w_U = \sqrt{2}/2 = 0.71$, indicating highest weight of urban features. With respect to the influence of L-features on non-weighted models discussed above (fig. 4f), local features seem to be of slightly lower importance. Meanwhile, U-features that are slightly anti-correlated with ${}_{s}\mathrm{R}_{\mathrm{avg}}^{2}$ in fig. 4g here exhibit relatively high importance. This may indicate that in the ${}_{s}\mathrm{R}_{\mathrm{avg}}^{2}$ metric for a given choice of feature set, weighting may fine-tune the impact of different categories.

These results suggest that optimised LB models outperform optimal CB models, and that weighted models outperform non-weighted models in the ${}_{s}\mathrm{R}_{\mathrm{avg}}^{2}$ metric and generally also in the other metrics applied. Differences between weighted and non-weighted models are small, and for both types of LB models as well as CB models, there is high negative bias and relatively low ${}_{s}\mathrm{RMSE}_{\mathrm{avg}}$ and ${}_{s}\mathrm{MAE}_{\mathrm{avg}}$. Nevertheless, the proposed feature selection methodology produces coherent results that may be used for feature set optimization. Such feature set optimization highlights the importance of local features and shows that weighting may act as a fine-tune adjustment method to inform on differential impacts of categories. Finally, weighting may have less impact for "better" models than for "worse" models because "better" models may already capture the underlying patterns sufficiently well.

## 3.2 Model Variance, Statistical Significance and Optimal Models

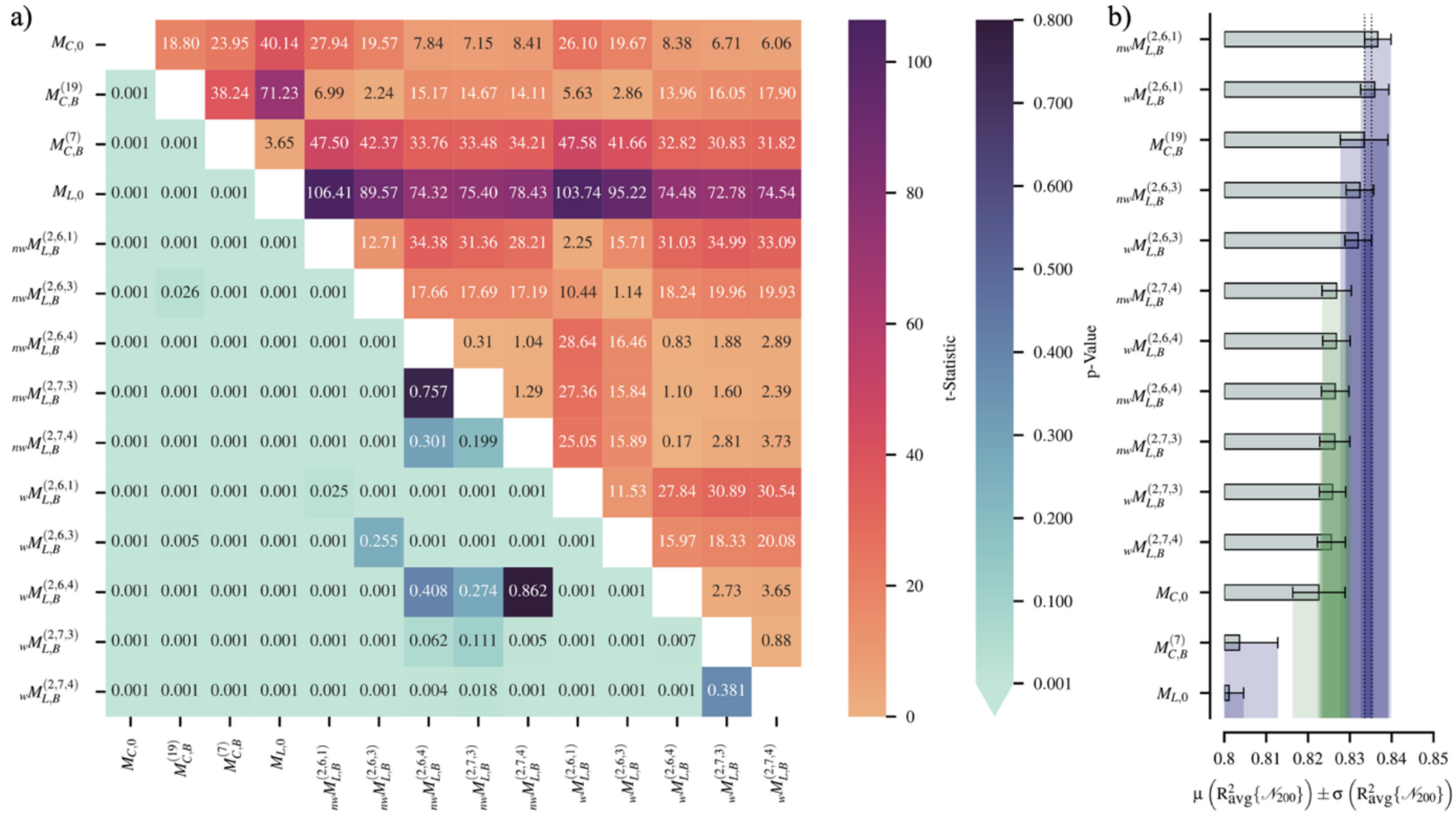

Figure 5: a: Upper triangular heatmap showing the $t$-statistics of pairwisely testing the models against the null hypothesis of identical means, lower triangular heatmap showing the pertaining $p$-values, displayed as 0.001 if they were smaller than 0.001 b: Accuracies of the models in descending order, with blue ribbons indicating standard deviations of the models, and the dashed area indicating the overlap of the $\mu\left(\mathrm{R}^2_{\mathrm{avg}}\{\mathcal{N}_{\epsilon_1=200}\}\right) \pm \sigma\left(\mathrm{R}^2_{\mathrm{avg}}\{\mathcal{N}_{\epsilon_1=200}\}\right)$ area of the top five models. The green ribbons and the other blue ribbons from the worst-performing models exhibit an overlap that is disjoint from the overlap of the top five models.

To inform on the expected intra-model variability, each model was ran $\epsilon_1 = 200$ times with different random data shuffling seeds to produce model distributions $\mathcal{N}_{\epsilon_1=200}$. Figure 5a shows the $t$-statistics and $p$-values of pairwise testing the means $\mu\left(\mathrm{R}^2_{\mathrm{avg}}\{\mathcal{N}_{\epsilon_1=200}\}\right)$ of the underlying distributions against the null hypothesis of being identical. $p$-values for testing the distributions against the null hypothesis of normal distribution (see table S3, supplementary materials), range between $0.0852 \leq p_N\left(\mathrm{R}^2_{\mathrm{avg}}\{\mathcal{N}_{\epsilon_1=200}\}\right) \leq 0.9975$. Under the assumed significance threshold of 0.005, this statistic thus suggests that all distributions are likely to very likely to be normal or near-normal. Likewise, only few pairs of models have $\mathcal{N}_{\epsilon_1=200}$-distributions whose means are not statistically significantly distinct. Where such statistically significant effects are present, the observed effects are therefore likely due to inclusion of different features and not due to variability within the models.

${}_{s}\mathrm{R}^2_{\mathrm{avg}}$ scores are generally not contained within the $\mu \pm \sigma$ range of pertaining $\mathcal{N}_{\epsilon_1=200}$-distributions (see text S6, supplementary materials). Nevertheless, within the five most accurate models ${}_{nw}M_{L,B}^{(2,6,1)}$, ${}_{w}M_{L,B}^{(2,6,1)}$, $M_{C,B}^{(19)}$, ${}_{nw}M_{L,B}^{(2,6,3)}$ and ${}_{w}M_{L,B}^{(2,6,3)}$ in order of descending $\mu\left(\mathrm{R}^2_{\mathrm{avg}}\{\mathcal{N}_{\epsilon_1=200}\}\right)$, the two top-performing LB models ${}_{nw}M_{L,B}^{(2,6,1)}$ and ${}_{w}M_{L,B}^{(2,6,1)}$ outperform the top performing CB model $M_{C,B}^{(19)}$. The differences in $\mu\left(\mathrm{R}^2_{\mathrm{avg}}\{\mathcal{N}_{\epsilon_1=200}\}\right)$ between these models are statistically significant ($p \leq 0.005$). After these five "top" models ($\mu\left(\mathrm{R}^2_{\mathrm{avg}}\{\mathcal{N}_{\epsilon_1=200}\}\right) = 0.83 \dots 0.84$, $\mu\left(\mathrm{RMSE}_{\mathrm{avg}}\{\mathcal{N}_{\epsilon_1=200}\}\right) = 2.18 \dots 2.21$ ˚C, $\mu\left(\mathrm{MAE}_{\mathrm{avg}}\{\mathcal{N}_{\epsilon_1=200}\}\right) = 1.65 \ldots 1.67\ ^{\circ}\mathrm{C}$, $\mu\left(\mathrm{MB}_{\mathrm{avg}}\{\mathcal{N}_{\epsilon_1=200}\}\right) = -1.20 \ldots -1.15\ ^{\circ}\mathrm{C}$) follow seven models of similar albeit lower accuracy. After these "intermediate" models follow two "worst" models — $M_{C,B}^{(7)}$ and $M_{L,0}$. "Top"-, "intermediate"- and "worst" models are distinct: any model contained in any accuracy group predicts statistically significantly different means $\mu\left(\mathrm{R}^2_{\mathrm{avg}}\{\mathcal{N}_{\epsilon_1=200}\}\right)$ than any model from any other group (cf. fig. 5b).

The "top" models are shown in fig. 6. In panels 6a-e, showing the difference between predicted and true target T2 values on 30.09.2019 15:00, differences across models are small, and differences between weighted and non-weighted models are very small. The strong negative bias of $\mu\left(\mathrm{MB}_{\mathrm{avg}}\{\mathcal{N}_{\epsilon_1=200}\}\right) = -1.20 \ldots -1.15\ ^{\circ}\mathrm{C}$ of the "top" models is visible during the day (figs. 6a-e). Conversely, T2 may be overestimated during the nighttime (figs. 6f-j, showing differences between predicted and true T2 values 12 hours later on 01.07.2019 03:00), likely resulting in a weak positive bias.

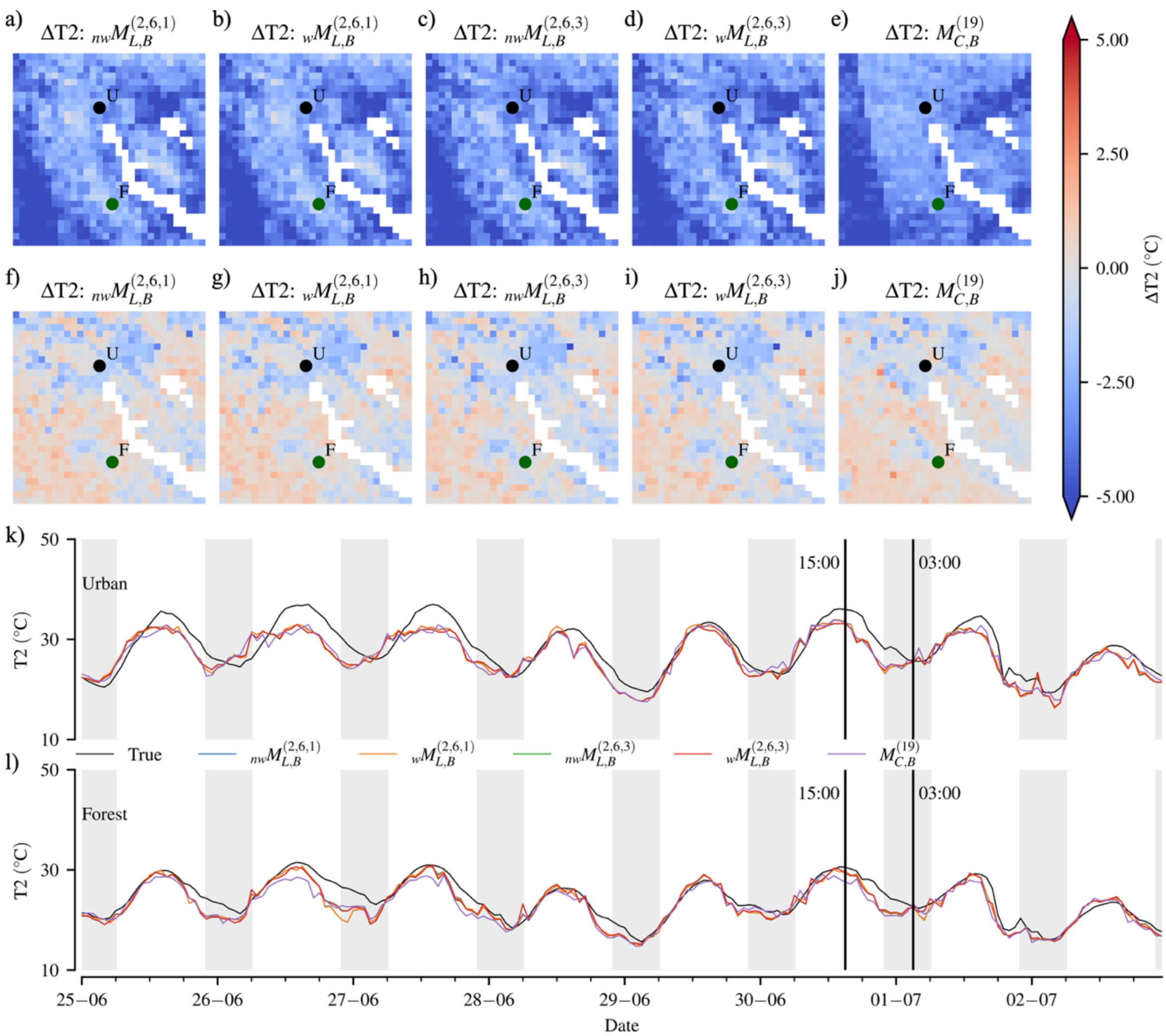

Figure 6: a – e: Differences between prediction and true values for T2 for the top 5 models at 30.06.2019 15:00. U and F indicate the same urban (LUT 13) and forest (LUT 5) location across all panels. Water areas (LUT 17) have been masked and appear white. f – j: Same as in a – e, but showing 01.07.2019 03:00. k – l: Timeseries for the testing period (HW week of 2019, 25.06.2019 00:00 until 02.07.2019 23:59) for the urban (LUT 13, panel k) and forest (LUT 5, panel l) location indicated in the panels above with U and F, respectively. Grey areas

indicate night (22:00 – 05:59). Black vertical lines indicate the two timepoints corresponding to the heatmaps (30.09.2019 15:00 in panels a – e, and 01.07.2019 03:00 in panels f – j).

The overall negative bias is also visible in the timeseries of the 2019 HW week (figs. 6kl): across all models, urban T2 is usually underestimated (fig. 6k), especially during the day. This is less accentuated for non-urban locations (fig. 6l), where all models manage to follow the true T2 evolution more closely. Similar behaviour was observed for TSK (see fig. S1, supplementary materials), although negative bias for TSK is lower than for T2.

### 3.2.1 Choice of Optimal Model

To assess mitigation potentials, high-performing ML models with high number of urban — or modifiable — features are desirable. When considering the "top" group, models including three U-features underperform models with a single U-feature. This gives rise to a trade-off: shall the model with highest accuracy or with highest number of U-features be picked?

It may be shown (see text S7 and fig. S2, supplementary materials) that the nominally "worst" of the "top" models with three urban features may be used instead of the "top" of the "top" models with only one urban feature, because associated accuracy losses (1) occur only *sometimes*; and (2) are likely small. Urban planners thereby gain two urban features — $M_{L,B}^{(2,6,1)}$ contains emissivity as the single U-feature whereas $M_{L,B}^{(2,6,3)}$ contains the three U-features emissivity (EMISS), albedo (ALBEDO) and leaf area index (LAI) — that they may use to mitigate UHIs.

Within these "top" of the "worst" of the "top" models — ${}_{nw}M_{L,B}^{(2,6,3)}$ and ${}_{w}M_{L,B}^{2,6,3}$ — shall the weighted or the non-weighted model be chosen to assess mitigation potentials? Although the results from weighting in the ${}_{s}\mathrm{R}^2_{\mathrm{avg}}$ metric and suggest that weighting improves performance, employing the $\mathrm{R}^2_{\mathrm{avg}}\{\mathcal{N}_{\epsilon_1=200}\}$ metric yields that weighting is likely statistically insignificant: observed differences in $\mu\left(\mathrm{R}^2_{\mathrm{avg}}\{\mathcal{N}_{\epsilon_1=200}\}\right)$ between weighted- and non-weighted LB models are associated with high $p$-values of $0.255$ for "worst" (${}_{nw}M_{L,B}^{(2,6,3)}$ and ${}_{w}M_{L,B}^{(2,6,3)}$) (very likely statistically insignificant), and $0.025$ for "top" (${}_{nw}M_{L,B}^{(2,6,1)}$ and ${}_{w}M_{L,B}^{(2,6,1)}$) (likely statistically insignificant) of the "top" models (fig. 5a). This is further supported by the very small differences between weighted and non-weighted models in fig. 6a-e. Likewise, within "intermediate" models, only one out of three configurations ($M_{L,B}^{(2,7,4)}$) shows statistically significant differences between weighted and non-weighted models. Therefore, even though weighted models outperform non-weighted models in the ${}_{s}\mathrm{R}^2_{\mathrm{avg}}$ metric, the weighted "worst" model shall here be disregarded in favour of the non-weighted "worst" model to assess mitigation potentials.

## 3.3 Impact of Number of Heat Wave Hours in Training Data

Having selected an optimal model — the non-weighted "worst" of the top models ${}_{nw}M_{L,B}^{(2,6,3)}$ that includes EMISS, ALBEDO and LAI as urban features — what is the optimal number of included heatwave data hours in the training set, and is this optimal number robust across other models? For the five "top" models, figures 7a-c and 7d-f show the evolution of ${}_{s}\mathrm{R}^2$ accuracies for including increasing 2017 heatwave data hours in the training set, always evaluated against the 2019 heatwave testing data, for the entire domain (fig. 7a-c) and for LUT 13 (urban- and built-up) only (fig. 7d-f), respectively. There are two general trends: first, there is a clear upwards trend of accuracy for both the entire domain and LUT 13 with increasing number of included heatwave hours. This trend is first rather small but increases towards the maximal number of included heatwave data. Secondly, for both the entire domain and LUT 13, the predictions in T2 contain more variability than the predictions for TSK (note the different axis scaling). The highest accuracies are universally observed when using the maximal hours of heatwave data. These results suggest that including more heatwave data increases the accuracy, both for T2 and TSK. Importantly, all predictions have been obtained using HPs tuned on the maximal number of 2017 HW data hours, and the largest increase in accuracy universally occurs from $144h$ to $168h$ in figs. 7a-f. Because the distributions of feature data between $144h$ and $168h$ of included HW data are supposedly similar, HP tuning *should* give similar HPs when tuned on $144h$ or $168h$. As it is within this range, however, that the largest increases in accuracy occur when HPs are kept the same, it is more probable that increases stem from inclusion of more HW data rather than from HP tuning effects.

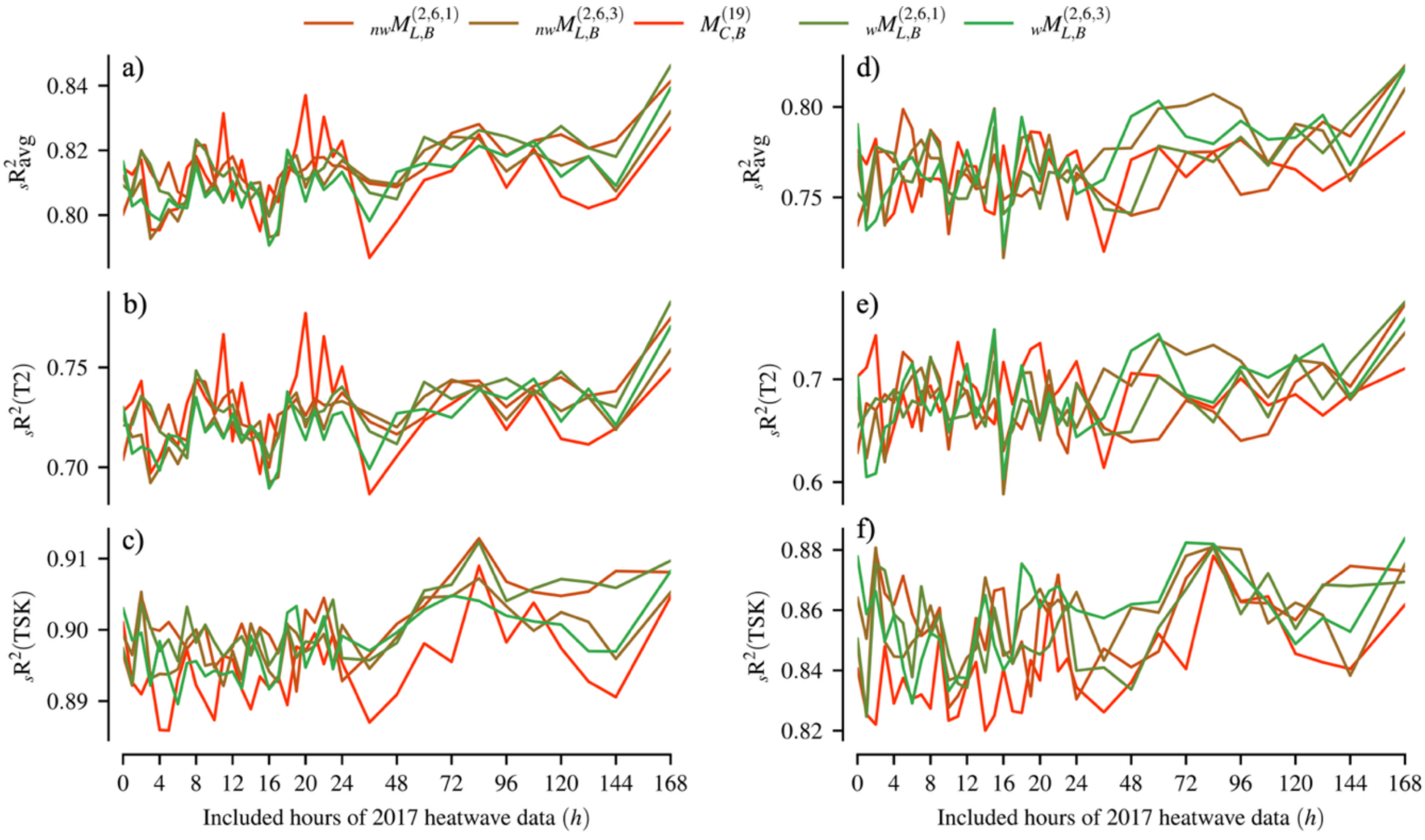

Figure 7: a-c: Including various hours of 2017 heatwave data in the training process, showing the combined- (a), T2- (b) and TSK-(c) accuracy, for the entire working domain. d-f: The same as in panels a – c, but only for LUT 13, urban and built-up.

### 3.4 Model Application to Estimate Mitigation Potentials

The non-weighted "worst" of the "top" models $_{nw}M_{L,B}^{(2,6,3)}$ was used to estimate the sensitivities of T2 and TSK towards variations in the included U-features emissivity, albedo and LAI, following the methodology in sec. 2.11. The imposed variations were chosen such that "meaningful" values resulted: for EMISS $\{{}^{\uparrow}\Delta_{\mathrm{EMISS}}, {}_{\downarrow}\Delta_{\mathrm{EMISS}}\} = \{-0.38, 0.013\}$ such that varied EMISS is within $[0.5, 0.998]$, for ALBEDO $\{{}^{\uparrow}\Delta_{\mathrm{ALBEDO}}, {}_{\downarrow}\Delta_{\mathrm{ALBEDO}}\} = \{-0.02, 0.25\}$ such that varied ALBEDO is within $[0.1, 0.982]$, and for LAI $\{{}^{\uparrow}\Delta_{\mathrm{LAI}}, {}_{\downarrow}\Delta_{\mathrm{LAI}}\} = \{-0.5, 5\}$ such that varied EMISS is within $[0.5, 6]$. Although in reality it is unfeasible to apply these changes homogeneously over the entire WD or to reach some of the tested values — ALBEDO values of 0.982 are very unlikely due to soling, for instance — the imposed variations may be used to construct a sensitivity estimate that may be robust for smaller imposed changes. The results are displayed in fig. 8 for T2 (see fig. S3, supplementary materials, for TSK). Gradient estimates are available in the supplementary materials (see table S6).

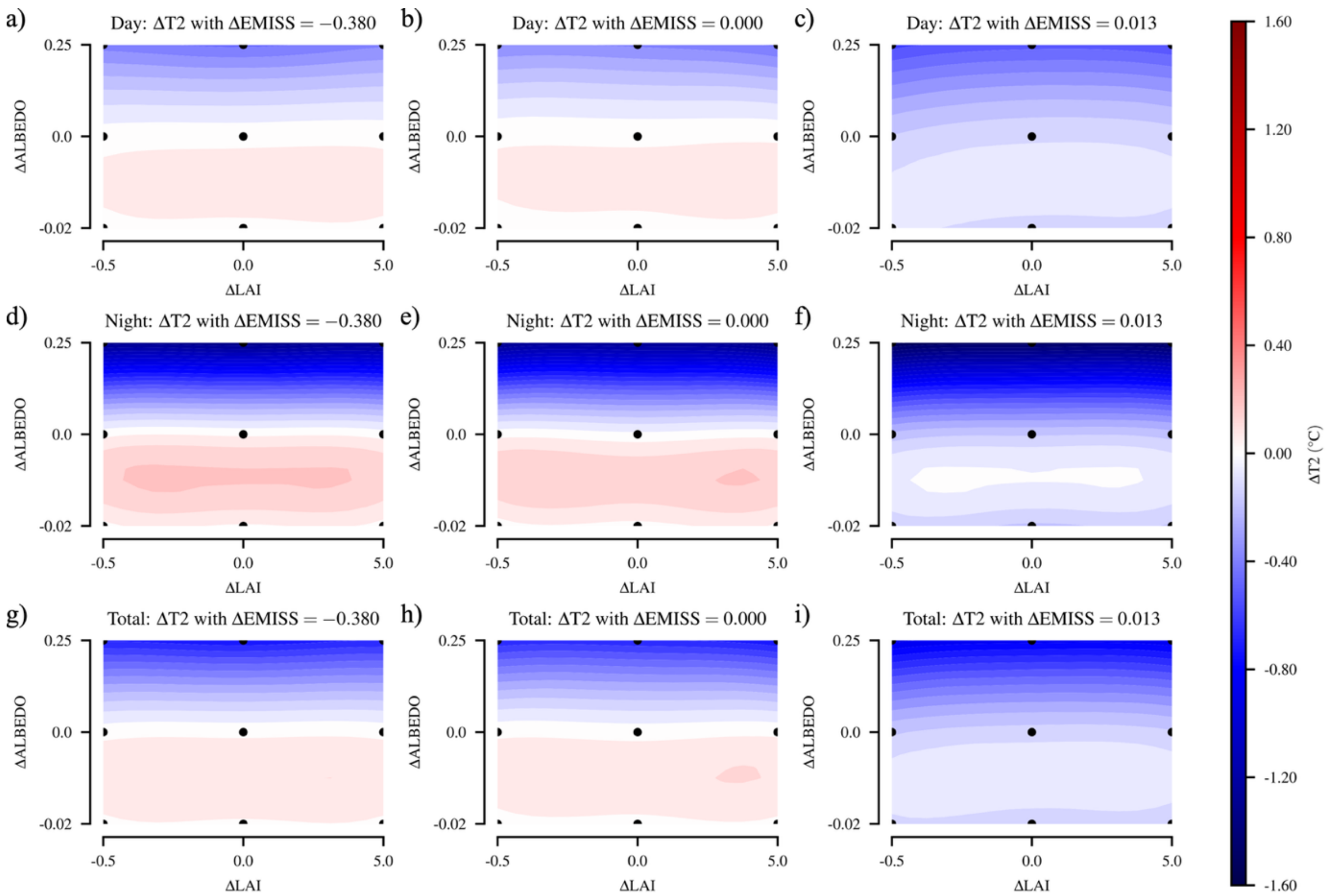

Figure 8: a-i: Heatmaps for ΔT2 resulting from applying the model $_{nw}M_{L,B}^{(2,6,3)}$ with a variable shift in the value of the top three urban features. Black dots indicate the points in the feature space where the model was evaluated, with the heatmap having been interpolated using cubic interpolation between the grid points. a-c: Response to variations during daytime (06:00 – 21:59). d-f: Response to variations during the nighttime (22:00 – 05:59). g-h: Total response to variations over the entire day.

For all EMISS values tested, and for the daytime-, nighttime- and total response, there is a strong dependency of the T2-response on ALBEDO variation, while the dependency on LAI appears weak (fig. 8a-i). Similar behaviour was observed for TSK (see fig. S3, supplementary materials), although variation with ALBEDO appears weaker and variations with LAI seem more pronounced. Furthermore, variations of T2 with ALBEDO appear lower during the day than during night, and EMISS variation response seem to have opposite effect for T2 and TSK.

Observed sensitivities (see table S6, supplementary materials) indicate similar results: the strongest cooling impact — or negative sensitivity — is observed for ALBEDO both for T2 and TSK, while EMISS responses differ in sign between T2 and TSK but are of largely similar magnitude. Effects of LAI are so low that uncertainties mostly do not permit determining the sign of the response, implying that response to LAI may be too low to be captured against the natural variability of the model used.

The magnitude of ALBEDO and EMISS response is up to 1 order of magnitude lower than computed in previous studies (Giannaros et al., 2018). Furthermore, as outlined in (Giannaros et al., 2018), higher emissivity should lead to lower T2 and lower TSK through higher radiative cooling, but while a strong cooling effect is observed for T2, TSK seems to increase. Together, these results suggest that mitigation strategies may lead to slightly different behaviour during the night, and that T2 and TSK may respond differently to the same mitigation strategy.

# 4 Discussion

The first aim of this study is to show that HW data is required to predict HW events, and that NHW data alone may be insufficient. This has been shown in fig. 7a-f, both globally and for urban- and built-up areas, with more HW data increasing accuracy in the ${}_{s}\mathrm{R}^2_{\mathrm{avg}}$ metric. The quality of the prediction is usually low, however: strong negative bias (negative ${}_{s}\mathrm{MB}_{\mathrm{avg}}$) and limited accuracy (relatively high ${}_{s}\mathrm{RMSE}_{\mathrm{avg}}$ and ${}_{s}\mathrm{MAE}_{\mathrm{avg}}$) are observable through all models. This may be a consequence of pre-processing of the data.

The second aim of this study is to demonstrate that LB models outperform CB models. The results obtained indicate that at least one optimised LB model outperforms the optimised CB model in both ${}_{s}\mathrm{R}^2_{\mathrm{avg}}$ and $\mu\left(\mathrm{R}^2_{\mathrm{avg}}\{\mathcal{N}_{\epsilon_1=200}\}\right)$ metrics, with ${}_{s}\mathrm{R}^2_{\mathrm{avg}}$ scores usually higher than the more robust $\mu\left(\mathrm{R}^2_{\mathrm{avg}}\{\mathcal{N}_{\epsilon_1=200}\}\right)$ scores but usually not contained within $\mu\left(\mathrm{R}^2_{\mathrm{avg}}\{\mathcal{N}_{\epsilon_1=200}\}\right) \pm \sigma\left(\mathrm{R}^2_{\mathrm{avg}}\{\mathcal{N}_{\epsilon_1=200}\}\right)$. Although this LB model was not the model finally chosen — a model including more urban features usable for mitigation strategy assessment than included in the best-performing model was chosen — it performs better than the optimised CB model in all other metrics, too, even though performance gains are small.

The third aim of this study is to demonstrate that LB models employing feature categorization and feature inclusion using FIOs lead to better results and higher interpretability. This has been demonstrated using $\mu\left(\mathrm{R}^2_{\mathrm{avg}}\{\mathcal{N}_{\epsilon_1=200}\}\right)$ scores: the findings of this study show that LB models employing this framework may outperform CB and LB models that do not employs the framework, emphasising that FIOs may contribute meaningful feature ranking schemes across scales. Finally, based on statistical arguments, an optimal model was selected to demonstrate how mitigation potentials of urban features may be estimated.

For this application, a non-weighted rather than a weighted model was selected, because weighting is likely statistically insignificant, or more precisely: within the $\mu\left(\mathrm{R}^2_{\mathrm{avg}}\{\mathcal{N}_{\epsilon_1=200}\}\right)$ metric, differences between weighted and non-weighted models were statistically insignificant. This means that within the $\mu\left(\mathrm{R}^2_{\mathrm{avg}}\{\mathcal{N}_{\epsilon_1=200}\}\right)$ metric, there is no need to adapt special weights for categories to obtain optimal performance and capture the nature of the system. Weights $w_D = w_L = w_U = 1$ as implied by non-weighted models are sufficient, because RFR-XGB in non-weighted mode appears robust in the $\mu\left(\mathrm{R}^2_{\mathrm{avg}}\{\mathcal{N}_{\epsilon_1=200}\}\right)$ metric. This robustness was hinted at also using ${}_{s}\mathrm{R}^2_{\mathrm{avg}}$ scores, where high-performing models were insensitive to different weights. The authors therefore recommend adapting $w_D = w_L = w_U = 1$.

As category weighting is a linear operation altering features values, feature space partition edges determining RFR-XGB predictions may be subjected to the same transformation, effectively rendering RFR-XGB weighting-invariant. Observed weighting effects in ${}_{s}\mathrm{R}^2_{\mathrm{avg}}$ imply that this invariance is likely only approximate. By averaging over large ensembles with different random shuffling seeds, slight changes in performance may be "averaged away" or become small against internal. Other machine learning methods could potentially increase the significance of weighting in the $\mu\left(\mathrm{R}^2_{\mathrm{avg}}\{\mathcal{N}_{\epsilon_1=200}\}\right)$ metric through resolving category weights explicitly instead of — indicated by the recommendation to use $w_D = w_L = w_U = 1$ — implicitly, although potentially at higher computational costs. RFR-XGB performance may likely be improved through other methodological adjustments: current preprocessing inhibits RFR-XGB to learn any relevant physical process on time scales exceeding data temporal resolution (i.e. hourly) and allows no spatial lateral information flow. However, synoptic-scale processes are associated with longer time scales, and build-up of heat in UAs may include memory effects as heat content is an integrated quantity. Additionally, small-scale processes such as heating or runoff may involve large lateral flows. These may be especially important for UAs because local features, shaped by the properties of both large- and small spatiotemporal scales, may be highly influential. Due to these interactions, assessment of spatial scale weights may be challenging when not including temporal evolution and lateral flow. Besides increasing weighting significance, these steps would likely render RFR-XGB much more accurate, which in turn may increase the confidence in estimates of mitigation potentials. Accounting for memory effects may be achieved through embedding past timesteps as features (Gudmundsson & Seneviratne, 2015) or performing CV through leaving out time blocks without prior random shuffling, thereby preserving the time series structures. Lateral information flow may be achieved through convolutional neuronal networks, that have the additional benefit of learning explicitly on features' values, thus circumventing the implicitness of weighting in this study.

All the same, the results obtained from applying the selected non-weighted model demonstrate mitigating effects of increased albedo, as proposed in literature (Giannaros et al., 2018). This emphasises the high interpretability of the proposed methodology in terms of mitigation strategy design. However, large-scale albedo increases (i.e. through using brighter construction materials) may appear easy-to-implement theoretically, but initially high albedo surfaces may be exposed to soiling (Shi et al., 2019), and high reflectance may induce other negative effects (Yang et al., 2015). Such effects were not included in this study, and they indicate that albedo increases as a mitigation strategy need to be carefully analysed. Unexpected increases of TSK when increasing EMISS and indeterminate sensitivity of TSK to LAI may also limit the confidence in mitigation potential assessments through the proposed framework. Shadowing and evaporative cooling effects of increased LAI were not explicitly modelled in WRF, potentially accounting for the indeterminate LAI sensitivity. Furthermore, if the underlying WRF model was forced at the land-atmosphere interface with remotely sensed radiological data for EMISS derived from TSK, this may lead to a tight functional relationship between EMISS and TSK. In that case imposing higher EMISS values may be misinterpreted by RFR-XGB as *higher* TSK at unchanged *effective* emissivity. This may indicate that the model partially overfits, and that increasing EMISS may still have a meaningful mitigating effect.

Overall, despite limitations, the proposed method gives clear results indicating feature importances and associated mitigation potentials. This makes the results straightforward to interpret for urban planners. To render the findings more robust, the method needs to be applied to more cities, because the results obtained in this study, and particularly the FIOs, are only applicable for the study WD of Zurich, and only for the present WRF output as training- and testing data. Other cities and other data configurations will very likely produce different results and different FIOs, and will require different pre-classifications. This is equally true when including more features, different data sources, or resolving urban 3D spaces explicitly. Lastly, urban microclimate models may be better suited to produce appropriate training data for RFR-XGB as mesoscale meteorological models such as WRF may overly smooth local effects (Carmeliet & Derome, 2024).

## 5 Conclusion

In this study, a machine learning emulator of Weather Research and Forecasting (WRF) model output was used to predict 2m-above ground- (T2) and surface temperature (TSK) for the city of Zurich, Switzerland. It proposes a novel method to assess the influence of features on different scales on urban heat islands (UHIs) and provides urban planners with a method to determine and quantify mitigation strategies for UHIs. This is achieved through pre-classifying features into groups pertaining to a certain scale and degree of modifiability, employing a feature selection scheme based on sequential backward selection (SBS) to create feature-inclusion orders (FIOs), and use feature set optimization to find optimal model configurations. Subsequently, the mitigation potential of the most modifiable and important features is determined. This study demonstrates (1) that random forest regression using extreme gradient boosting (RFR-XGB) to predict heat waves (HWs) in urban areas (UAs) performs better if more HW data is included in the training set; (2) that land-based (LB) models distinguishing land use types (LUTs) may outperform city-based (CB) models not distinguishing LUTs; and (3) that LB models employing feature categorization and feature selection using LUT-specific FIOs may maximise performance and interpretability.

The results of this study indicate that the proposed methodology may be applicable to characterise UHIs, and to assess the potential of mitigation strategies. The most important urban features identified in this study are emissivity, albedo and leaf area index (LAI). Signs of estimated sensitivities of T2 towards these are in line with previous studies, even though different response magnitudes are observed. Sensitivities of TSK towards emissivity, and to a lesser degree LAI, may contradict previous studies and require further investigation. The results also indicate that weighting of categories likely does not statistically significantly change results compared to no weighting of categories. Along with required evidence of robustness across different cities, the method needs to be improved with respect to methodological adjustments in subsequent studies. The results of this study should be understood to indicate that the proposed method may yield meaningful, highly interpretable results for the city it is applied to.

It should be emphasised that the methodology is intended to be applicable for any UA, although it has been demonstrated only for Zurich, where results demonstrate the general applicability of the method. Future work should focus on testing the method on other cities to assess its efficiency and robustness on a larger set of UAs. Moreover, future work should focus on improving the methodology by using time-step embedding and convolutional neuronal networks, or other methods to include lateral flow and reenforce weighting impact, and address overfitting to reduce the respective insignificances and uncertainties associated with weighting and mitigation potential estimates.

## 6 Acknowledgements

D. Tschan acknowledges supervision through Y. Zhao, discussion on WRF configuration with D. Strebel and helpful insights from Z. Wang and J. Carmeliet.

## 7 Conflict of Interests

The authors declare that no there are no conflicts of interest.

## 8 Data Availability Statement

Code for this study is freely available under https://gitlab.com/uhi3/fsf_project_2.git.

Supporting Information for

**Interpretable Machine Learning for Urban Heat Mitigation: Attribution and Weighting of Multi-Scale Drivers**

David Immanuel Tschan[1,2], Zhi Wang[1], Dominik Strebel[1], Jan Carmeliet[1], Yongling Zhao[1]

[1]Department of Mechanical and Process Engineering, ETH Zürich, Switzerland

[2]Department of Environmental Systems Science, ETH Zürich, Switzerland

## Contents of this file

## Additional Supporting Information (Files uploaded separately)

Captions for Tables S1, S3 and S5

## Introduction

The following documents contains detailed methodology and supplementary analysis required to interpret the results. It furthermore presents supplementary information on input data and results in tables. Lastly, three additional figures are included.

## Text S1. Justification of Feature Categories

Driving features are solar zenith angle (COSZEN), terrain height (HGT), soil type (ISLTYP), rain mixing ratio (QRAIN), orographic variance (VAR) and subgrid orographic variance (VAR_SSO). Clearly, solar zenith angle and terrain heigh are not modifiable by urban planners, as is the underlying soil type, even though UAs may partially or fully remove topsoil or render it impermeable through urban morphology. More difficult to assess is perhaps rain mixing ratio: this depends on temperature (rain evaporation), humidity, cloud cover and pressure. It is unlikely that any mitigation strategy might thus specifically target rain mixing ratio without feeding back onto these other parameters. Easier are (sub-grid-) orographic variance, because topography and orography are not in human control.

Local features include foliage canopy water content (CANWAT), longwave radiation at bottom (LWDNB), planetary boundary layer height (PBLH), surface pressure (PSFC), surface runoff (SFROFF), shortwave radiation at bottom (SWDNB), soil temperature at bottom (TMN), zonal wind 10 m above ground (U10), underground runoff (UDROFF) and meridional wind 10 m above ground (V10). Canopy water content (CANWAT) may partially be controlled through the type of vegetation introduced, but background climate and geographic location, i.e. being in an arid or a humid climate, will greatly influence what kind of vegetation may be sustained. Longwave (LWDNB) and short wave radiation (SWDNB) is subject to the canopy content, itself subject to this geographic constraint. Although canopy design may alter incoming and outgoing long- and shortwave radiation, geographic location will still be dominating in the radiation balance. The planetary boundary layer height (PBLH) is determined by atmospheric stability and the vertical potential temperature gradient. These are influenced by synoptic-scale weather, but also through urban-scale humidity and temperature modifications. Similarly, the pressure (PSFC) is coupled to wind- (U10, V10) and temperature fields, again being somewhere in between large-scale and small-scale processes. Through construction and soil type, soil temperature (TMN) is largely controlled, but background climate will greatly matter here. Similarly, runoff (SFROFF, UDROFF) is determined partially by soil permeability, but also by precipitation and evapotranspiration which are large-scale processes eluding human control.

Urban features include albedo (ALBEDO), emissivity (EMISS), vegetation type (IVGTYP), leaf area index (LAI), and vegetation fraction (VEGFRA). Albedo (ALBEDO) and emissivity (EMISS) may be controlled by choosing appropriate surface materials, although soiling effects will change the reflective or emissive properties over time (Shi et al., 2019). Vegetation type (IVGTYP) and vegetation amount (VEGFRA, LAI) may be determined by urban planners upon plantation.

**Text S2. Detailed Description of Metrics used**

Let $y_i$, $\hat{y}_i$ and $\bar{y}_i$ denote the true-, predicted- and predicted-average target values from the testing process, for a total number of $N$ samples. Then:

- The $\mathrm{R}^2 \in (-\infty, 1]$ score, or coefficient of determination, measures the degree of observed variance explicable through the model. As more variance becomes explicable, $\mathrm{R}^2$ increases. Formally, it is defined as:

$$\mathrm{R}^2 = \frac{\sum_i^N (y_i - \hat{y}_i)^2}{\sum_i^N (y_i - \bar{y}_i)^2} \tag{S1}$$

- The root mean squared error $\mathrm{RMSE} \in [0, \infty)$ is a measure of accuracy that decreases as the model fits the data better. It is formally defined as:

$$\mathrm{RMSE} = \sqrt{\frac{1}{N} \sum_i^N (y_i - \hat{y}_i)^2} \tag{S2}$$

The RMSE score may be interpreted as the average distance between prediction and true values. This metric has the same unit as the target values.

- The mean absolute error MAE $\in [0, \infty)$ is formally defined as:

$$\text{MSE} = \frac{1}{N}\sum_i^N |y_i - \hat{y}_i| \quad \text{(S3)}$$

It may be interpreted as the average offset between predicted and true values.

- The mean bias score MB $\in (-\infty, \infty)$ is formally defined as:

$$\text{MB} = \frac{1}{N}\sum_i^N (y_i - \hat{y}_i) \quad \text{(S4)}$$

It may be interpreted as the average difference between predicted and true values. This average difference is often referred to as bias.

Generally, a prediction becomes better if $R^2$ increases, and RMSE and MAE decrease, and if the mean bias MB moves closer to 0.

**Text S3. Detailed Methodology for Feature Ranking Scheme**

Let $F^{(n)}$ denote the initial feature set consisting of $n$ features, i.e. $F^{(n)} = \{f_1, f_2, \dots, f_n\}$. A feature ranking may be implemented as follows:

1. Initialise a list ${}_*F$.
2. Using the feature set $F^{(n-t)}$ at iteration $t$, fit $n-t$ models on the $n-t$ feature subsets $F^{(n-t)} \setminus f_k = \{f_i\}_{i \neq k}^{n-t}$. This results in a set of $n-t$ scores $\{r_i\}_{i=1}^{n-t} = \left\{\mathrm{R}^2_{\mathrm{avg}}(\{f_i\}_{i \neq k}^{n-t})\right\}_{k=1}^{n-t}$.
3. Determine $1 \leq j \leq n-t$ such that $r_j = \min_k \{r_k\}_{k=1}^{n-t}$, which pertains to feature $f_j$.
4. Set $F^{(n-[t+1])} = F^{(n-t)} \setminus f_j$.
5. Determine the number of features in $F^{(n-[t+1])}$.
    a. If $F^{(n-[t+1])}$ contains more than one feature, append $f_j$ to ${}_*F$, set $t \leftarrow t+1$ and go to 2.
    b. If $F^{(n-[t+1])}$ contains one feature, append this feature to ${}_*F$ and terminate.

Thus in each iteration, the feature selection scheme removes the feature whose removal causes the *highest* loss in $\mathrm{R}^2_{\mathrm{avg}}$ score relative to the removal of all other features, and appends this feature the FIO ${}_*F$. After termination ${}_*F = \left\{{}_*f_i\right\}_{i=1}^{n}$ constitutes a feature set containing the initial number $n$ of features *ordered* — indicated through the subscript $*$ — in descending importance, where for $1 \leq j < n$ the feature ${}_*f_j$ is more important, in a hierarchical sense (cf. main text sec. 2.7), than ${}_*f_{j+1}$. Such an ordered feature set ${}_*F$ is referred to as a feature inclusion order (FIO).

## Text S4. Detailed Methodology for Weighting of Categories

As weighting occurs after scaling, all weights $w_{D,L,U}$ are constrained to $w_{D,L,U} \in [0, 1]$ and to $w_{D,L,U} \geq 0$. Assuming the respective ratios of the weights, rather than their absolute values, to be relevant, the weight vector may be further constrained to $\|\boldsymbol{w}\| = 1$. With this, the weight space may be parametrized using $\phi, \theta$-space:

$$w_D = \sin\theta \cos\phi \,, w_L = \sin\theta \sin\phi \,, w_U = \cos\theta \tag{S5}$$

with $0 \leq \theta, \phi \leq \frac{\pi}{2}$. For any given LUT, and for any given tuple of number of included features $({}_{*}n_D,\ {}_{*}n_L,\ {}_{*}n_U)$, the differential impact of different categories onto the result is reflected in assigning different weights to categories. This can be seen as a fine-tuning of the non-weighted method to clarify the contribution of different categories to the total outcome. Finding optimal weights amounts to finding optimal $\theta, \phi$. This was implemented in this study using grid search by choosing the best combination of weights resulting from setting $\theta_i, \phi_j$ to any point within the search grid $\theta_\alpha = \alpha\Delta\theta$ and $\phi_\alpha = \alpha\Delta\phi$ with $0 \leq \alpha \leq 10$ and $\Delta\theta = \Delta\phi = 0.05\pi$ (i.e. an $11 \times 11$ search grid).

## Text S5. Detailed Methodology for Model Application for Mitigation Potential Estimation

Mitigation potentials associated with the ${}_{*}n_U$ U-features over UAs (LUT 13) may be estimated for each U-feature ${}_{*}f_k$ within $F^{({}_{*}n_U)} = \{{}_{*}f_k\}_{k=1}^{{}_{*}n_U}$ by in- or decreasing the testing data of that feature by ${}^{\uparrow}\Delta_k$ or ${}_{\downarrow}\Delta_k$, respectively, from its unvaried value where $\Delta_k = 0$ is assumed. The effect of altering U-feature ${}_{*}f_k$ may then be measured by fitting the pertaining model on the unvaried feature data and testing on the varied feature that. The method allows both to alter single U-features, or several simultaneously.

The change in the target value $\Delta\widehat{\mathrm{T}}_i$ for model $i$ and target T — either T2 or TSK — is then estimated as the difference between average predictions from non-varied and varied U-features. For this, the predictions from the non-varied distribution $\widehat{\mathrm{T}}(\mathcal{N}_{i,\epsilon_1=200})$ and the varied distribution $\widehat{\mathrm{T}}(\mathcal{N}_{i,\epsilon_2=200})$ for the $i$-th model are used. In the ensemble of $\epsilon_2 = 200$ models, different random seeds were used to train on unvaried data and test against varied data. First, the responses are grouped by hours of the day:

$$\left\{\left\{\widehat{\mathrm{T}}_{i,h}\right\}_z^S\right\}_h^H = \left\{\bigcup_j^{\epsilon_1}\left(\widehat{\mathrm{T}}\left\{\mathcal{N}_{i,j} \middle| \text{hour} = h\right\}\right)\right\}_h^H \tag{S6a}$$

$$\left\{\left\{\widehat{\mathrm{T}}'_{i,h}\right\}_z^S\right\}_h^H = \left\{\bigcup_j^{\epsilon_2}\left(\widehat{\mathrm{T}}\left\{\mathcal{N}_{i,j} \middle| \text{hour} = h\right\}\right)\right\}_h^H \tag{S6b}$$

Each hourly slice $\left\{\widehat{\mathrm{T}}_{i,h}\right\}_z^S$ and $\left\{\widehat{\mathrm{T}}'_{i,h}\right\}_z^S$ contains $S$ data points that result from aggregating the response over LUT 13. For any given hour $h$, the average response $\mu\left(\widehat{\mathrm{T}}_{i,h}\right)$ or $\mu\left(\widehat{\mathrm{T}}'_{i,h}\right)$ may then be computed as

$$\mu\left(\widehat{\mathrm{T}}_{i,h}\right) = \frac{1}{S}\sum_{z}^{S}\left\{\widehat{\mathrm{T}}_{i,h}\right\}_{z}^{S} \tag{S7a}$$

$$\mu\left(\widehat{\mathrm{T}}'_{i,h}\right) = \frac{1}{S}\sum_{z}^{S}\left\{\widehat{\mathrm{T}}'_{i,h}\right\}_{z}^{S} \tag{S7b}$$

For any given period of the day — day, night or the entire total day — the average response for that period $\mu\left(\widehat{\mathrm{T}}_i\right)$ or $\mu\left(\widehat{\mathrm{T}}'_i\right)$ can be found by averaging the hourly average responses over that period. Supposing the period to consist of the hours $\{h_P\}_p^P$, the period response becomes

$$\mu\left(\widehat{\mathrm{T}}_{\mathrm{i}}\right) = \frac{1}{p}\sum_{p}^{P}\mu\left(\widehat{\mathrm{T}}_{i,h_p}\right) \tag{S8a}$$

$$\mu\left(\widehat{\mathrm{T}}'_i\right) = \frac{1}{p}\sum_{p}^{P}\mu\left(\widehat{\mathrm{T}}'_{i,h_p}\right) \tag{S8b}$$

So that finally $\Delta\widehat{\mathrm{T}}_i$ may be estimated:

$$\Delta\widehat{\mathrm{T}}_i = \mu\left(\widehat{\mathrm{T}}'_i\right) - \mu\left(\widehat{\mathrm{T}}_i\right) \tag{S9}$$

Union of the predicted responses in eqs. (S6a) and (S6b) occurs only over LUT 13 (urban- and built-up), and over the entire diurnal cycle. To obtain estimates on the uncertainty of the change in target value, $u\left(\Delta\widehat{\mathrm{T}}_i\right)$, the standard errors associated with the entire data, $\mathrm{se}\left(\widehat{\mathrm{T}}_i\right)$ and $\mathrm{se}\left(\widehat{\mathrm{T}}'_i\right)$, are computed and added for the observed period:

$$\mathrm{u}\left(\Delta\widehat{T}_i\right) = \mathrm{se}\left(\widehat{\mathrm{T}}_i\right) + \mathrm{se}\left(\widehat{\mathrm{T}}'_i\right) \tag{S10}$$

$$\mathrm{se}\left(\widehat{\mathrm{T}}_{i,h}\right) = \frac{\sigma\left(\bigcup_{p}^{P}\left\{\widehat{\mathrm{T}}_{i,h_p}\right\}_{z}^{S_p}\right)}{\sqrt{\sum_{h}^{H} S_p}} \tag{S11}$$

$$\mathrm{se}\left(\widehat{\mathrm{T}}'_{i,h}\right) = \frac{\sigma\left(\bigcup_{p}^{P}\left\{\widehat{\mathrm{T}}'_{i,h}\right\}_{z}^{S_p}\right)}{\sqrt{\sum_{h}^{H} S_p}} \tag{S12}$$

where $S_p$ indicates the number of data points in the hourly slice $h_p$. Finally, the target sensitivity $\Delta\widehat{\mathrm{T}}_i\ \delta^{-1}_{{}_*f_k}$ towards unit change of ${}_*f_k$, that is $\delta^{-1}_{{}_*f_k}$, may be estimated via

$$\Delta\widehat{\mathrm{T}}_\imath\ \delta^{-1}_{{}_*f_k} = \frac{\partial\Delta\widehat{\mathrm{T}}_i}{\partial\ {}_*f_k} \approx \frac{\Delta\widehat{\mathrm{T}}_i\left({}_*f_k + {}^{\uparrow}\Delta_k\right) - \Delta\widehat{\mathrm{T}}_i\left({}_*f_k - {}_{\downarrow}\Delta_k\right)}{{}^{\uparrow}\Delta_k - {}_{\downarrow}\Delta_k} \tag{S13}$$

The associated uncertainty of the gradient estimate, $\mathrm{u}\left(\Delta\widehat{\mathrm{T}}_\imath\ \delta^{-1}_{{}_*f_k}\right)$ is

$$\mathrm{u}\left(\Delta\widehat{\mathrm{T}}_\imath\ \delta^{-1}_{{}_*f_k}\right) = \frac{\mathrm{u}\left(\Delta\widehat{\mathrm{T}}_i\left(+\ {}^{\uparrow}\Delta_k\right)\right) + \mathrm{u}\left(\Delta\widehat{\mathrm{T}}_i\left(-\ {}_{\downarrow}\Delta_k\right)\right)}{{}^{\uparrow}\Delta_k - {}_{\downarrow}\Delta_k} \tag{S14}$$

This allows to assess the mitigation potential of $_{*}f_k$, permitting to design mitigation strategies against UHIs based on expected temperature change associated with changes in feature $_{*}f_k$.

**Text S6. Deviation between $s$-Scores and Ensemble Scores**

The absolute difference between distribution mean accuracy and $_{s}\mathrm{R}^2_{\mathrm{avg}}$ score, $\Delta\mathrm{R}^2 = \left|\mu\left(\mathrm{R}^2_{\mathrm{avg}}\{\mathcal{N}_{\epsilon_1=200}\}\right) - {}_{s}\mathrm{R}^2_{\mathrm{avg}}\right|$ is larger than the standard deviation, i.e. $\Delta\mathrm{R}^2 > \sigma\left(\mathrm{R}^2_{\mathrm{avg}}\{\mathcal{N}_{\epsilon_1=200}\}\right)$, for all models but $M^{(19)}_{C,B}$, $M^{(7)}_{C,B}$, $_{nw}M^{(2,6,3)}_{L,B}$ and $_{nw}M^{(2,7,4)}_{L,B}$. The largest difference pertains to $_{w}M^{(2,6,1)}_{L,B}$ and is $\Delta\mathrm{R}^2 = 3.049\sigma\left(\mathrm{R}^2_{\mathrm{avg}}\{\mathcal{N}_{\epsilon_1=200}\}\right)$, with mean and median deviations across all models of $\Delta\mathrm{R}^2 = 1.5308\sigma$ and $\Delta\mathrm{R}^2 = 1.5702\sigma$, respectively. Assuming normally distributed $\mathrm{R}^2_{\mathrm{avg}}\{\mathcal{N}_{\epsilon_1=200}\}$ — justified by $p_N\left(\mathrm{R}^2_{\mathrm{avg}}\{\mathcal{N}_{\epsilon_1=200}\}\right)$ values — it can be expected that 68% of $_{s}\mathrm{R}^2_{\mathrm{avg}}$ scores lie within $\left(\mathrm{R}^2_{\mathrm{avg}}\{\mathcal{N}_{\epsilon_1=200}\}\right) \pm \sigma(\mathrm{R}^2_{\mathrm{avg}}\{\mathcal{N}_{\epsilon_1=200}\})$. The observed value is only 28.6%, implying that $_{s}\mathrm{R}^2_{\mathrm{avg}}$ scores are generally not contained in the $\mu \pm \sigma$ range of the associated distributions.

**Text S7. Statistical Argument for Model Choice**

Figure S1 shows the accuracy distributions of all models considered. The "top" models are indicated (fig. S1e shows best model, indicated by "1" in the title, followed by fig. S1j, indicated by "2" in the title, and so on). Note that "top" models indicate also a probability of observing a score within the overlap, indicated as a blue ribbon. These probabilities range between 11.5% (fig. S1b) and 17.6% (fig. S1j).

As can be seen from fig. 5b, $\mu\left(\mathrm{R}^2_{\mathrm{avg}}\{\mathcal{N}_{\epsilon_1=200}\}\right) \pm \sigma(\mathrm{R}^2_{\mathrm{avg}}\{\mathcal{N}_{\epsilon_1=200}\})$ of the "top" models overlap between $0.8335 \leq \mu(\mathrm{R}^2_{\mathrm{avg}}\{\mathcal{N}_{\epsilon_1=200}\}) \leq 0.8352$. Assuming normal distributions of the "top" models — justified by $p_N(\mathrm{R}^2_{\mathrm{avg}}\{\mathcal{N}_{\epsilon_1=200}\})$ in table S3 — the probability of any top-five model to produce a score within the overlap, $P_i(O)$, ranges between 11.5% and 17.6% (cf. fig. S1). Once in the overlap, the probability that the score can be attributed to the $i$-th model is

$$P(O \wedge M_i) = \frac{P_i(O)}{\sum_i P_i(O)} \qquad \text{(S15)}$$

$P(O \wedge M_i)$ is maximal for $_{w}M^{(2,6,1)}_{L,B}$ at 22.6% and minimal for $M^{19}_{C,B}$ at 14.8%. This means that a score in the overlap may in 77.4% of cases have been produced by the "best" of the "top" models ($_{nw}M^{(2,6,1)}_{L,B}$ and $_{w}M^{(2,6,1)}_{L,B}$) *or* the "worst" of the "top" models ($_{nw}M^{(2,6,3)}_{L,B}$ or $_{w}M^{2,6,3}_{L,B}$). The maximal expected accuracy loss between "best" and "worst" of the "top" models (cf. table S3) is $\Delta\mu\left(\mathrm{R}^2_{\mathrm{avg}}\{\mathcal{N}_{\epsilon_1=200}\}\right) = 0.0047 \pm 0.0064$ (likely statistically significant as "top" and "worst" of the "top" models are statistically significantly different at $p = 0.001$) when assuming $\sigma$ to be an appropriate estimation of the model uncertainty — which is perhaps not certain given the unexpected behaviour of $\Delta\mathrm{R}^2$ (see sec. S6). This loss is only slightly bigger than $\sigma(\mathrm{R}^2_{\mathrm{avg}}\{\mathcal{N}_{\epsilon_1=200}\})$ of "top" and “worst” of the “top” models — and may therefore have been produced through the internal variability of the model.

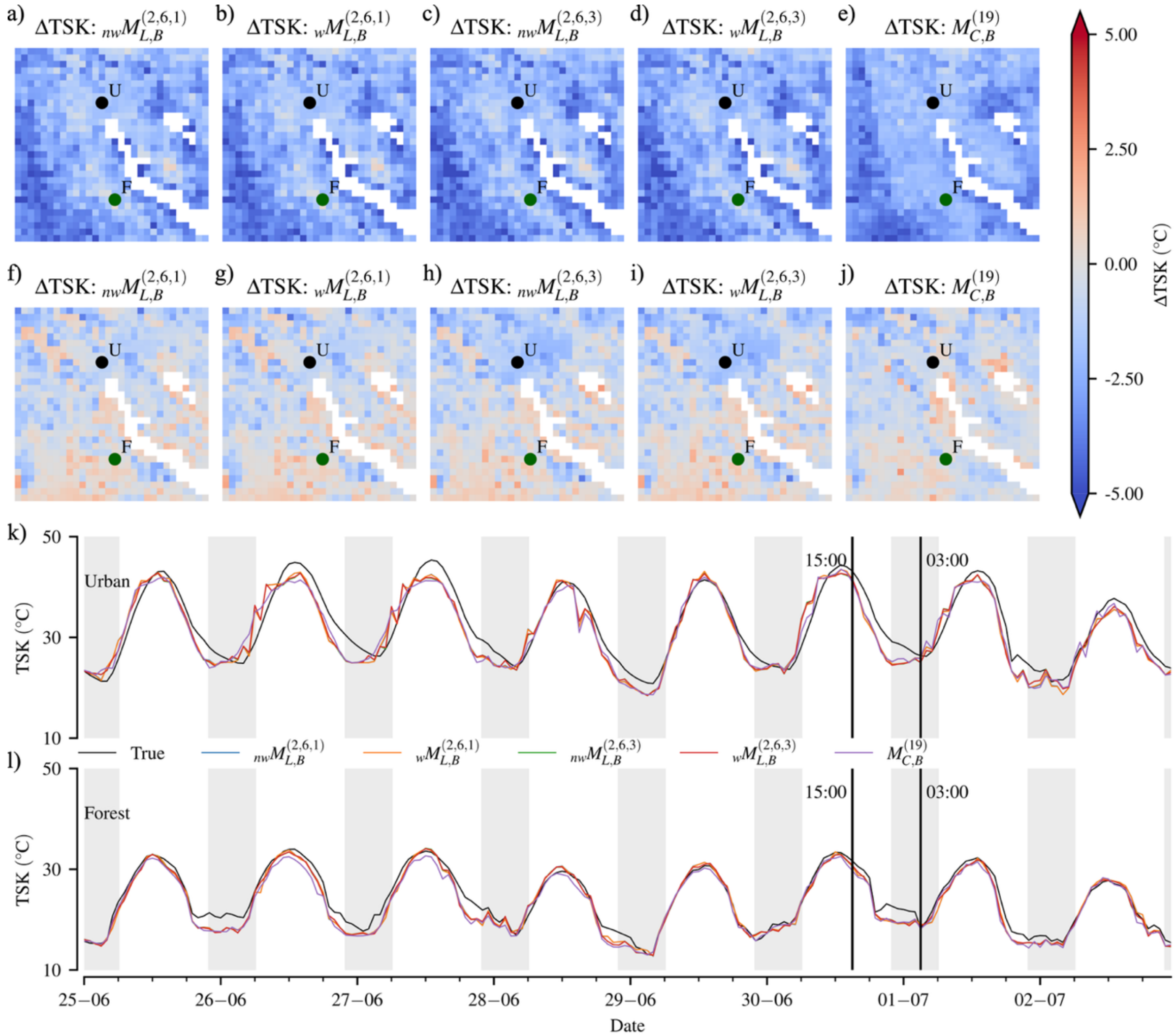

Figure S1. The same as in fig. 6, but for TSK.

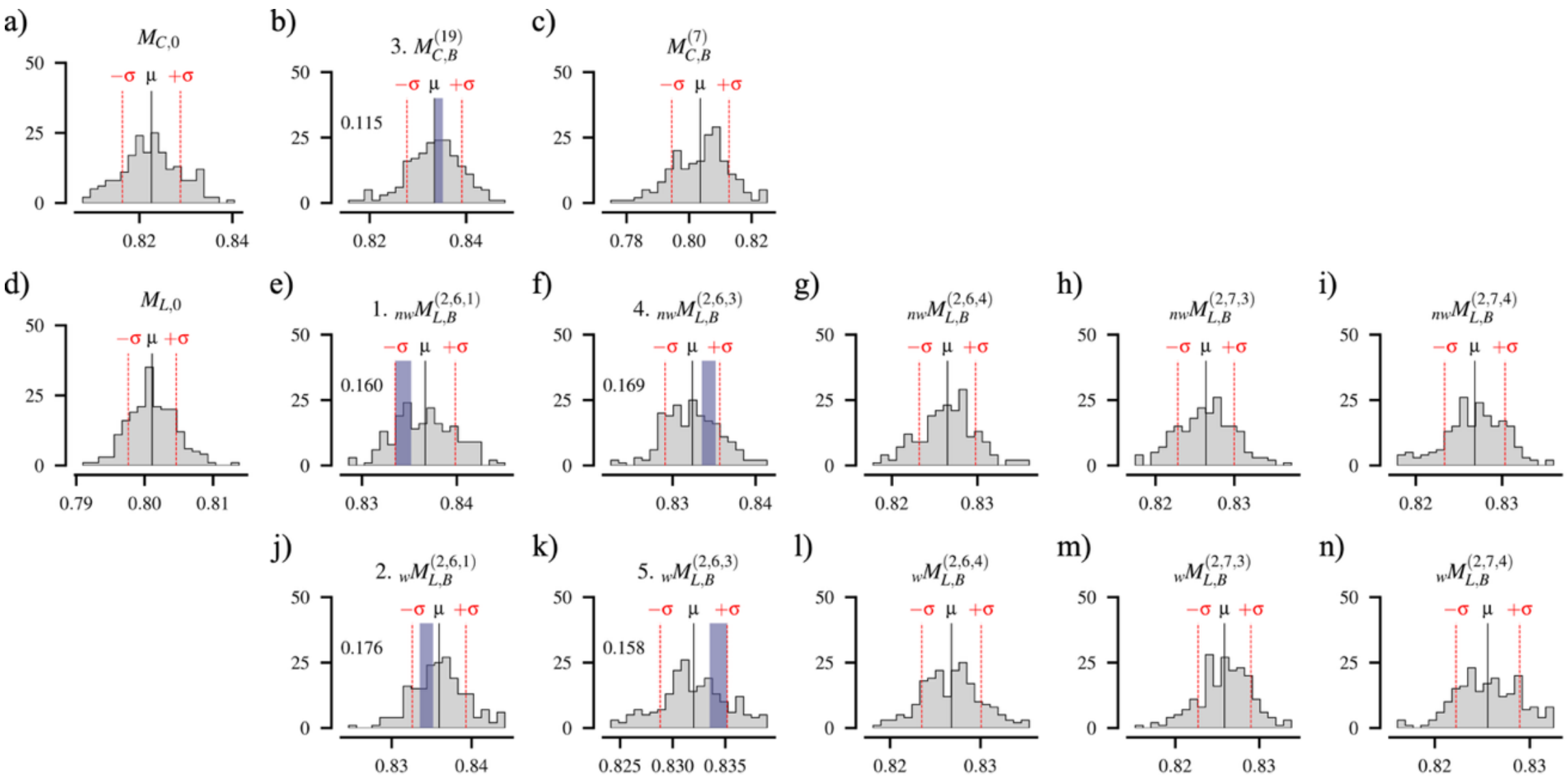

Figure S2. Histograms of the model accuracy distributions $\left(\mathrm{R}^2_{\mathrm{avg}}\{\mathcal{N}_{\epsilon_1=200}\}\right)$ for all models considered. The five "top" models are indicated in the plot titles with performance ranks 1-5 marked in the respective titles. These "top" models also show a blue band that indicates the overlap in $0.8335 \leq \mathrm{R}^2_{\mathrm{avg}}\{\mathcal{N}_{\epsilon_1=200}\} \leq 0.8352$ established in text S7. The plots where the overlapping region (identified in fig. 5b) is indicated also show the probability of finding an accuracy score within the overlap when fitting a normal distribution probability density on the pertaining histograms. a-c: CB models. d-n: LB models.

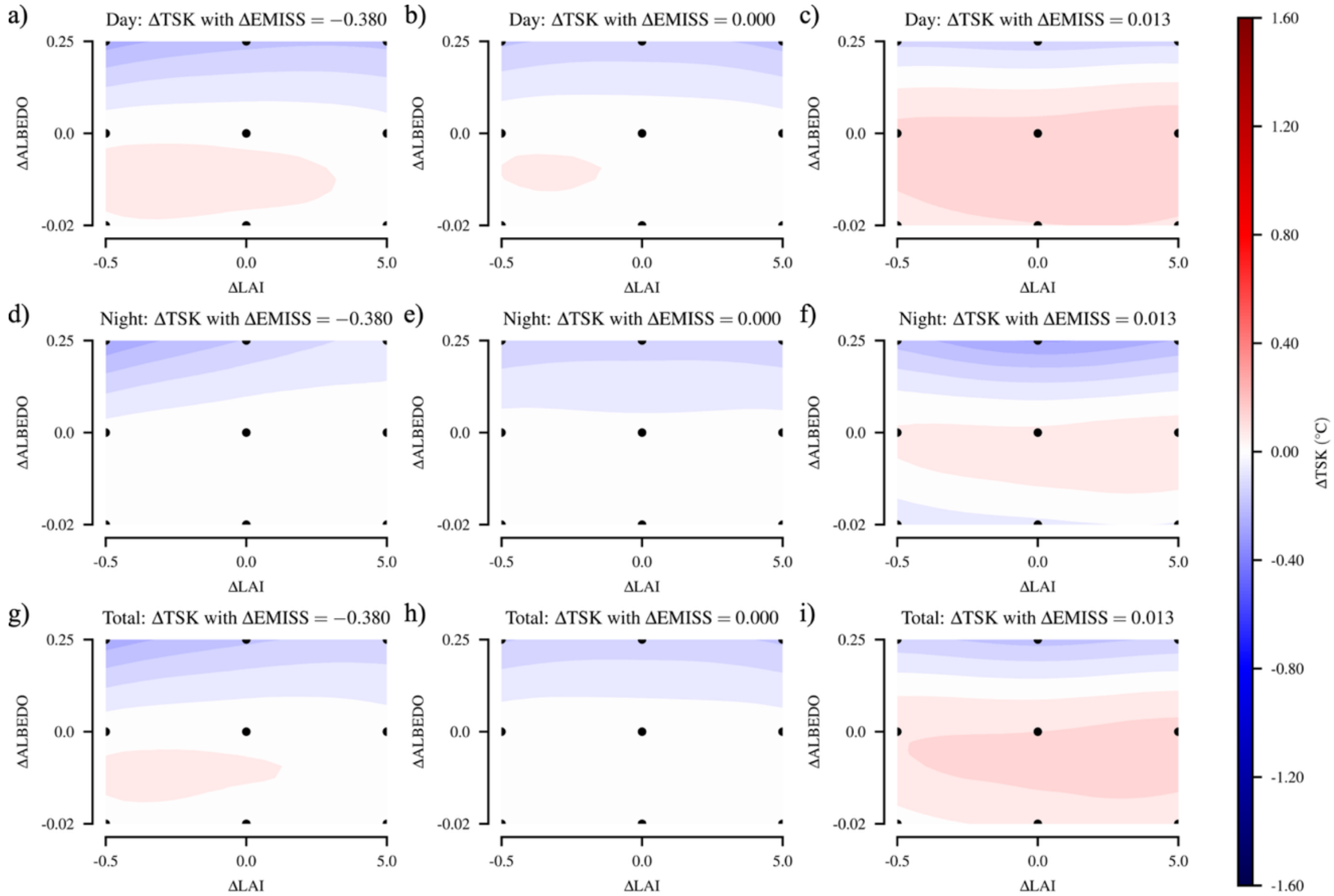

Figure S3. The same as in fig. 8, but for TSK.

Table S1. Features and Targets obtained from the WRF-SLUCM output for 2017, 2019 and combined 2017 + 2019. The presented data does not include LUT 17 (water). Shown are mean $\mu$, standard deviation $\sigma$ and skewness $sk$ for 2017 data, 2019 data and combined 2017 + 2019 data. Footnotes: a: driving features; b: local features; c: urban features

| LUT Type | LUT Description |
|---|---|
| **1** | **Evergreen Needleleaf Forest** |
| **2** | **Evergreen Broadlead Forest** |
| 3 | Decidious Needleleaf Forest |
| **4** | **Decidious Broadleaf Forest** |
| **5** | **Mixed Forests** |
| 6 | Closed Shrublands |
| **7** | **Open Shrublands** |
| **8** | **Woody Savannas** |
| 9 | Savannas |
| **10** | **Grasslands** |
| **11** | **Permanent Wetlands** |
| **12** | **Croplands** |
| **13** | **Urban and Built-Up** |
| **14** | **Cropland/Natural Vegetation Mosaic** |
| 15 | Snow and Ice |
| 16 | Barren or Sparsely Vegetated |
| 17 | Water |

Table S2. Overview over the LUTs used in this study. Bold entries highlight LUTs present in the WD. LUT classification followed the classification for MODIS land cover products from Sulla-Menashe & Friedl, 2018.

Table S3. Overview over performance of models in all metrics used. Listed models include baseline models, optimised CB, as well as top non-weighted- and weighted LB models. The weights $w_D$, $w_L$, $w_U$ pertaining to D-, L- and U-features for weighted LB models are reported, along with scores in ${}_{s}\mathrm{X}_{\mathrm{avg}}$- (identical random shuffling seed) and $\mu\left(\mathrm{X}_{\mathrm{avg}}\{\mathcal{N}_{\epsilon_1}\}\right)$-metrics (ensemble). Always, $\epsilon_1 = 100$. Additionally, the standard deviation $\sigma\left(\mathrm{X}_{\mathrm{avg}}\{\mathcal{N}_{\epsilon_1}\}\right)$ and the $p$-value of testing the distribution against the null hypothesis of being normally distributed, $\mathrm{p}_N\left(\mathrm{R}^2_{\mathrm{avg}}\{\mathcal{N}_{\epsilon_1=200}\}\right)$, are reported. Fields containing "—" indicate that the column value is not available for the pertaining model.

| Rank | 1 | 2 | 3 | 4 | 5 | 6 | 7 | 8 | 9 | 10 |
|---|---|---|---|---|---|---|---|---|---|---|
| Feature | LWDNB | PBLH | CANWAT | UDROFF | SWDNB | ALBEDO | EMISS | SFROFF | V10 | U10 |
| Rank | 11 | 12 | 13 | 14 | 15 | 16 | 17 | 18 | 19 | 20 |
| Feature | QRAIN | TMN | ISLTYP | VAR_SSO | VAR | LU_INDEX | IVGTYP | VEGFRA | COSZEN | HGT |
| Rank | 21 | 22 | | | | | | | | |
| Feature | PSFC | LAI | | | | | | | | |

Table S4. Feature-inclusion order (FIO) for the CB model

Table S5. Feature-inclusion order (FIO) for the LB model.

| Period | $\Delta\widehat{\mathrm{T}}\ \delta_{\mathrm{EMISS}}^{-1}$ ( °C) | $\Delta\widehat{\mathrm{T}}\ \delta_{\mathrm{ALBEDO}}^{-1}$ ( °C) | $\Delta\widehat{\mathrm{T}}\ \delta_{\mathrm{LAI}}^{-1}$ ( °C) |
|---|---|---|---|
| *T2* | | | |
| Day | $-0.3018 \pm 0.0184$ | $-1.4105 \pm 0.0277$ | $0.0010 \pm 0.0013$ |
| Night | $-0.4904 \pm 0.0202$ | $-4.7893 \pm 0.0363$ | $-0.0005 \pm 0.0014$ |
| Total | $-0.3647 \pm 0.0183$ | $-2.5368 \pm 0.0284$ | $0.0005 \pm 0.0013$ |
| *TSK* | | | |
| Day | $0.2727 \pm 0.0316$ | $-0.5527 \pm 0.0466$ | $-0.0028 \pm 0.0023$ |
| Night | $0.0779 \pm 0.0203$ | $-0.4146 \pm 0.0280$ | $-0.0006 \pm 0.0013$ |
| Total | $0.2078 \pm 0.0310$ | $-0.5067 \pm 0.0453$ | $-0.0020 \pm 0.0022$ |

Table S6. Overview over model application results, showing variation in target prediction $\Delta\widehat{\mathrm{T}}$ per unit feature change $\delta_f$.

# Table S1

| Name | Description | Unit | 2017 | | | 2019 | | | 2019 + 2017 | | |
|---|---|---|---|---|---|---|---|---|---|---|---|
| | | | $\mu$ | $\sigma$ | $sk$ | $\mu$ | $\sigma$ | $sk$ | $\mu$ | $\sigma$ | $sk$ |
| *Features* | | | | | | | | | | | |
| ALBEDO[c] | Surface Albedo | — | 0.1675 | 0.0236 | 0.0990 | 0.1768 | 0.0515 | 4.6555 | 0.2769 | 0.1184 | 1.3864 |
| CANWAT[b] | Foliage Canopy Water Content | $kgm^{-2}$ | 0.0161 | 0.0716 | 5.1394 | 0.1080 | 0.1785 | 1.4126 | 0.1412 | 0.1689 | 0.9974 |
| COSZEN[a] | Zenith angle | rad | 0.2895 | 0.4407 | 0.0084 | 0.2881 | 0.4403 | 2.2058e-5 | 0.2669 | 0.4347 | 0.0757 |
| EMISS[c] | Emissivity | — | 0.9469 | 0.0349 | -1.0805 | 0.9470 | 0.0346 | -1.0945 | 0.9542 | 0.0220 | -1.7413 |
| HGT[a] | Terrain Height | m | 520.8086 | 88.7170 | 0.6643 | 520.8084 | 88.7170 | 0.6443 | 520.8086 | 88.7170 | 0.6643 |
| ISLTYP[a] | Soil Type | — | | | | | *Categorical* | | | | |
| IVGTYP[c] | Vegetation Type | — | | | | | *Categorical* | | | | |
| LAI[c] | Leaf Area Index | $m^2m^{-2}$ | 4.1214 | 1.8139 | -0.6957 | 4.1307 | 1.8066 | -0.7254 | 4.7378 | 1.1546 | -0.8191 |
| LU_INDEX | Land Use Type | — | | | | | *Categorical* | | | | |
| LWDNB[b] | Longwave Radiation at Bottom | $Wm^{-2}$ | 365.0426 | 26.8786 | -0.1434 | 349.0592 | 29.9627 | -0.1395 | 351.3814 | 30.6930 | -0.1449 |
| PBLH[b] | Planetary Bounary Layer Height | m | 351.4272 | 410.1020 | 1.2524 | 333.9043 | 372.0216 | 1.3355 | 470.1146 | 376.6531 | 1.0780 |
| PSFC[b] | Surface Pressure | Pa | 95977.5100 | 989.8087 | -0.5968 | 95750.8800 | 1068.3694 | -0.5169 | 95612.9000 | 1219.2892 | -0.2974 |
| QRAIN[a] | Rain Mixing Ratio | $kgkg^{-1}$ | 6.127E-7 | 2.0485E-5 | 61.1704 | 6.3888e-4 | 6.2328e-5 | 23.9743 | 4.4418e-5 | 1.5931e-4 | 9.2420 |
| SFROFF[b] | Surface Runoff | mm | 1.5562 | 2.0549 | 1.6003 | 13.8998 | 13.5140 | 1.3147 | 8.4936 | 11.6579 | 2.0135 |
| SWDNB[b] | Shortwave Radiation at Bottom | $Wm^{-2}$ | 351.0178 | 366.3360 | 0.4579 | 325.3036 | 355.6135 | 0.5846 | 488.9915 | 332.0749 | -0.0433 |
| TMN[b] | Soil Temperature at Bottom | °C | 281.9639 | 0.5637 | -0.5788 | 218.964 | 0.5637 | -0.5791 | 281.9627 | 0.5666 | -0.5772 |
| U10[b] | Wind Speed ($10m$ above ground, u-direction) | $ms^{-1}$ | 0.3573 | 2.6807 | -0.5871 | -0.2331 | 2.5251 | 0.1457 | -0.1389 | 2.5593 | 0.0239 |
| UDROFF[b] | Underground Runoff | mm | 30.2999 | 8.0227 | -0.4292 | 147.6104 | 42.4052 | -0.8758 | 127.6706 | 58.7390 | -0.5682 |
| V10[b] | Wind Speed ($10m$ above ground, v-direction) | $ms^{-1}$ | -0.6268 | 2.0953 | -0.1590 | -0.9402 | 2.0878 | 0.3270 | -0.8847 | 2.0932 | 0.2457 |
| VAR[a] | Orographic Variance | — | 106.6717 | 37.5529 | 2.0111 | 106.6717 | 37.5529 | 2.0111 | 115.7756 | 48.4775 | 1.2886 |
| VAR_SSO[a] | Sub-Grid Orographic Variance | — | 10669.8650 | 5269.0640 | 1.7840 | 10669.8640 | 5269.0635 | 1.7840 | 10669.8640 | 5269.0635 | 1.7840 |
| VEGFRA[c] | Vegetation Fraction | — | 73.1027 | 14.1765 | -2.2747 | 73.0550 | 14.0191 | -2.3906 | 73.5866 | 13.0102 | -2.1425 |
| *Targets* | | | | | | | | | | | |
| T2 | Air Temperature ($2m$ above ground) | °C | 23.8338 | 4.4820 | -0.3103 | 20.2205 | 5.7111 | 0.0136 | 20.8521 | 6.4341 | -0.1437 |
| TSK | Surface Temperature | °C | 24.3853 | 6.7107 | 0.1850 | 20.8254 | 7.5817 | 0.1886 | 21.6946 | 8.2821 | 0.0707 |

# Table S3

| Model | Weights | | | s-Metrics | | | | Ensemble Metrics | | | | | Normality |
|---|---|---|---|---|---|---|---|---|---|---|---|---|---|
| | $w_D$ | $w_L$ | $w_U$ | ${}_s\mathrm{R}^2_{\mathrm{avg}}$ | ${}_s\mathrm{RMSE}_{\mathrm{avg}}$ | ${}_s\mathrm{MAE}_{\mathrm{avg}}$ | ${}_s\mathrm{MB}_{\mathrm{avg}}$ | $\mu\left(\mathrm{R}^2_{\mathrm{avg}}\{\mathcal{N}_{\epsilon_1}\}\right)$ | $\sigma\left(\mathrm{R}^2_{\mathrm{avg}}\{\mathcal{N}_{\epsilon_1}\}\right)$ | $\mu\left(\mathrm{RMSE}_{\mathrm{avg}}\{\mathcal{N}_{\epsilon_1}\}\right)$ | $\mu\left(\mathrm{MAE}_{\mathrm{avg}}\{\mathcal{N}_{\epsilon_1}\}\right)$ | $\mu\left(\mathrm{MB}_{\mathrm{avg}}\{\mathcal{N}_{\epsilon_1}\}\right)$ | $\mathrm{p}_{\mathcal{N}}\left(\mathrm{R}^2_{\mathrm{avg}}\{\mathcal{N}_{\epsilon_1=200}\}\right)$ |
| *Baseline Models* | | | | | | | | | | | | | |
| $M_{C,0}$ | — | — | — | 0.8163 | 2.2836 | 1.7552 | -1.3016 | 0.8226 | 0.006253 | 2.2492 | 1.7191 | -1.2832 | 0.5086 |
| $M_{L,0}$ | 1 | 1 | 1 | 0.8084 | 2.3641 | 1.8295 | -1.4519 | 0.8011 | 0.003496 | 2.3834 | 1.8330 | -1.4595 | 0.2065 |
| *Optimised CB Models* | | | | | | | | | | | | | |
| $M_{c,B}^{(19)}$ | — | — | — | 0.8302 | 2.2406 | 1.7022 | -1.2202 | 0.8334 | 0.005692 | 2.1897 | 1.6647 | -1.1940 | 0.0852 |
| $M_{C,B}^{(7)}$ | — | — | — | 0.8096 | 2.4619 | 1.7065 | -0.7213 | 0.8036 | 0.009162 | 2.3705 | 1.6717 | -0.9072 | 0.2699 |
| *Optimised LB Models (non-weighted)* | | | | | | | | | | | | | |
| ${}_{nw}M_{L,B}^{(2,6,1)}$ | 1 | 1 | 1 | 0.8414 | 2.1603 | 1.6424 | -1.1371 | 0.8367 | 0.003161 | 2.1786 | 1.6544 | -1.1466 | 0.3400 |
| ${}_{nw}M_{L,B}^{(2,6,3)}$ | 1 | 1 | 1 | 0.8321 | 2.2145 | 1.6721 | -1.2124 | 0.8324 | 0.003289 | 2.2092 | 1.6699 | -1.2029 | 0.8394 |
| ${}_{nw}M_{L,B}^{(2,6,4)}$ | 1 | 1 | 1 | 0.8319 | 2.2047 | 1.6686 | -1.2617 | 0.8265 | 0.003288 | 2.2420 | 1.7024 | -1.2674 | 0.8602 |
| ${}_{nw}M_{L,B}^{(2,7,3)}$ | 1 | 1 | 1 | 0.8302 | 2.2208 | 1.6911 | -1.2834 | 0.8264 | 0.003581 | 2.2432 | 1.7036 | -1.2865 | 0.9975 |
| ${}_{nw}M_{L,B}^{(2,7,4)}$ | 1 | 1 | 1 | 0.8302 | 2.2123 | 1.6799 | -1.1979 | 0.8268 | 0.003499 | 2.2287 | 1.6955 | -1.2353 | 0.1833 |
| *Optimised LB Models (weighted)* | | | | | | | | | | | | | |
| ${}_{w}M_{L,B}^{(2,6,1)}$ | 0.9755 | 0.1545 | 0.1564 | 0.8462 | 2.1312 | 1.6186 | -1.1043 | 0.8359 | 0.003369 | 2.1830 | 1.6569 | -1.1504 | 0.7567 |
| ${}_{w}M_{L,B}^{(2,6,3)}$ | 0.9045 | 0.2939 | 0.3090 | 0.8393 | 2.1699 | 1.6378 | -1.1802 | 0.8320 | 0.003191 | 2.2111 | 1.6718 | -1.2037 | 0.7066 |
| ${}_{w}M_{L,B}^{(2,6,4)}$ | 0.0000 | 0.8090 | 0.5878 | 0.8332 | 2.2037 | 1.6728 | -1.2518 | 0.8268 | 0.003283 | 2.2404 | 1.7013 | -1.2675 | 0.8589 |
| ${}_{w}M_{L,B}^{(2,7,3)}$ | 0.4156 | 0.4156 | 0.8090 | 0.8328 | 2.2188 | 1.6902 | -1.2695 | 0.8259 | 0.003140 | 2.2462 | 1.7068 | -1.2877 | 0.1300 |
| ${}_{w}M_{L,B}^{(2,7,4)}$ | 0.3673 | 0.7208 | 0.5878 | 0.8335 | 2.1985 | 1.6738 | -1.2200 | 0.8256 | 0.003347 | 2.2354 | 1.7001 | -1.2458 | 0.5748 |

# Table S5

| LUT | Driving | | Local | | Urban | |
|---|---|---|---|---|---|---|
| | Rank | Feature | Rank | Feature | Rank | Feature |
| 1 | 1 | COSZEN | 1 | LWDNB | 1 | ALBEDO |
| Evergreen | 2 | QRAIN | 2 | PBLH | 2 | IVGTYP |
| Needleleaf Forest | 3 | VAR | 3 | CANWAT | 3 | EMISS |
| | 4 | VAR_SSO | 4 | UDROFF | 4 | LAI |
| | 5 | HGT | 5 | SFROFF | 5 | VEGFRA |
| | 6 | ISLTYP | 6 | SWDNB | | |
| | | | 7 | V10 | | |
| | | | 8 | U10 | | |
| | | | 9 | TMN | | |
| | | | 10 | PSFC | | |
| 2 | 1 | COSZEN | 1 | LWDNB | 1 | ALBEDO |
| Evergreen | 2 | QRAIN | 2 | PBLH | 2 | IVGTYP |
| Broadleaf Forest | 3 | VAR_SSO | 3 | CANWAT | 3 | EMISS |
| | 4 | HGT | 4 | SWDNB | 4 | LAI |
| | 5 | VAR | 5 | PSFC | 5 | VEGFRA |
| | 6 | ISLTYP | 6 | UDROFF | | |
| | | | 7 | TMN | | |
| | | | 8 | SFROFF | | |
| | | | 9 | U10 | | |
| | | | 10 | V10 | | |
| 4 | 1 | COSZEN | 1 | LWDNB | 1 | EMISS |
| Decidious | 2 | QRAIN | 2 | PBLH | 2 | ALBEDO |
| Broadleaf Forest | 3 | VAR_SSO | 3 | CANWAT | 3 | IVGTYP |
| | 4 | HGT | 4 | UDROFF | 4 | VEGFRA |
| | 5 | VAR | 5 | SFROFF | 5 | LAI |
| | 6 | ISLTYP | 6 | SWDNB | | |
| | | | 7 | V10 | | |
| | | | 8 | U10 | | |
| | | | 9 | PSFC | | |
| | | | 10 | TMN | | |
| 5 | 1 | COSZEN | 1 | PBLH | 1 | EMISS |
| Mixed Forests | 2 | QRAIN | 2 | LWDNB | 2 | ALBEDO |
| | 3 | VAR_SSO | 3 | CANWAT | 3 | IVGTYP |
| | 4 | HGT | 4 | UDROFF | 4 | VEGFRA |
| | 5 | VAR | 5 | SFROFF | 5 | LAI |
| | 6 | ISLTYP | 6 | SWDNB | | |
| | | | 7 | V10 | | |
| | | | 8 | PSFC | | |
| | | | 9 | U10 | | |
| | | | 10 | TMN | | |
| 7 | 1 | COSZEN | 1 | LWDNB | 1 | ALBEDO |
| Open Shrublands | 2 | QRAIN | 2 | PBLH | 2 | EMISS |
| | 3 | HGT | 3 | CANWAT | 3 | IVGTYP |
| | 4 | VAR_SSO | 4 | UDROFF | 4 | VEGFRA |
| | 5 | VAR | 5 | SFROFF | 5 | LAI |
| | 6 | ISLTYP | 6 | SWDNB | | |
| | | | 7 | U10 | | |
| | | | 8 | V10 | | |
| | | | 9 | TMN | | |
| | | | 10 | PSFC | | |
| 8 | 1 | COSZEN | 1 | LWDNB | 1 | ALBEDO |
| Woody Savannas | 2 | QRAIN | 2 | SFROFF | 2 | EMISS |
| | 3 | HGT | 3 | CANWAT | 3 | IVGTYP |
| | 4 | VAR | 4 | PBLH | 4 | LAI |
| | 5 | VAR_SSO | 5 | SWDNB | 5 | VEGFRA |
| | 6 | ISLTYP | 6 | U10 | | |
| | | | 7 | TMN | | |
| | | | 8 | UDROFF | | |
| | | | 9 | V10 | | |
| | | | 10 | PSFC | | |
| 10 | 1 | QRAIN | 1 | LWDNB | 1 | LAI |
| Grasslands | 2 | COSZEN | 2 | PBLH | 2 | ALBEDO |
| | 3 | HGT | 3 | CANWAT | 3 | EMISS |
| | 4 | VAR_SSO | 4 | UDROFF | 4 | IVGTYP |
| | 5 | VAR | 5 | SFROFF | 5 | VEGFRA |
| | 6 | ISLTYP | 6 | SWDNB | | |
| | | | 7 | V10 | | |
| | | | 8 | U10 | | |
| | | | 9 | TMN | | |
| | | | 10 | PSFC | | |
| 11 | 1 | COSZEN | 1 | SWDNB | 1 | ALBEDO |
| Permanent | 2 | QRAIN | 2 | CANWAT | 2 | IVGTYP |
| Wetlands | 3 | HGT | 3 | SFROFF | 3 | LAI |
| | 4 | VAR | 4 | LWDNB | 4 | EMISS |
| | 5 | VAR_SSO | 5 | PBLH | 5 | VEGFRA |
| | 6 | ISLTYP | 6 | UDROFF | | |
| | | | 7 | TMN | | |
| | | | 8 | V10 | | |
| | | | 9 | U10 | | |
| | | | 10 | PSFC | | |
| 12 | 1 | COSZEN | 1 | SWDNB | 1 | EMISS |
| Croplands | 2 | QRAIN | 2 | PBLH | 2 | ALBEDO |
| | 3 | HGT | 3 | CANWAT | 3 | IVGTYP |
| | 4 | VAR_SSO | 4 | UDROFF | 4 | LAI |
| | 5 | VAR | 5 | PSFC | 5 | VEGFRA |
| | 6 | ISLTYP | 6 | LWDNB | | |
| | | | 7 | SFROFF | | |
| | | | 8 | U10 | | |
| | | | 9 | V10 | | |
| | | | 10 | TMN | | |
| 13 | 1 | QRAIN | 1 | SWDNB | 1 | EMISS |
| Urban and Built- | 2 | COSZEN | 2 | PBLH | 2 | ALBEDO |
| Up | 3 | HGT | 3 | PSFC | 3 | LAI |
| | 4 | VAR_SSO | 4 | CANWAT | 4 | IVGTYP |
| | 5 | VAR | 5 | UDROFF | 5 | VEGFRA |
| | 6 | ISLTYP | 6 | LWDNB | | |
| | | | 7 | U10 | | |
| | | | 8 | SFROFF | | |
| | | | 9 | V10 | | |
| | | | 10 | TMN | | |
| 14 | 1 | COSZEN | 1 | LWDNB | 1 | ALBEDO |
| Cropland/Natural | 2 | QRAIN | 2 | PBLH | 2 | EMISS |
| Vegetation Mosaic | 3 | HGT | 3 | CANWAT | 3 | IVGTYP |
| | 4 | VAR_SSO | 4 | UDROFF | 4 | VEGFRA |
| | 5 | VAR | 5 | SFROFF | 5 | LAI |
| | 6 | ISLTYP | 6 | SWDNB | | |
| | | | 7 | V10 | | |
| | | | 8 | U10 | | |
| | | | 9 | PSFC | | |
| | | | 10 | TMN | | |